\documentclass[twocolumn,traditabstract]{aa}
\usepackage{fixltx2e}
\usepackage[english]{babel}
\usepackage{graphicx,amsmath}
\usepackage{epsf,color}
\usepackage[mathscr]{eucal}
\usepackage{amsmath}
\usepackage{amssymb,amsfonts}
\usepackage{natbib}
\usepackage{txfonts}
\usepackage{dsfont}
\let\tmp\newinsert
\let\newinsert\newbox
\usepackage{floatrow}
\let\newinsert\tmp
\usepackage{color}

\usepackage{hyperref}

\usepackage{caption}
\usepackage{slantsc}
\usepackage{subfig}

\newcommand{\TTH}{{ {\scshape The Three Hundred}}}
\newcommand{\HRH}{{\bf {\scshape hr\_hydro}}}
\newcommand{\LRH}{{\bf {\scshape lr\_hydro}}}
\newcommand{\HRD}{{\bf {\scshape hr\_dmonly}}}
\newcommand{\LRD}{{\bf {\scshape lr\_dmonly}}}
\newcommand{\BMONE}{\bf {\scshape MB1 }}
\newcommand{\BMTWO}{\bf{\scshape MB2 }}
\newcommand{\BMTHREE}{\bf {\scshape MB3 }}
\newcommand{\BMFOUR}{\bf {\scshape MB4 }}
\newcommand{\Euclid}{{Euclid }}

\begin{document}
\title{\textsc{the three hundred} : contrasting clusters galaxy density in hydrodynamical and dark matter simulations}

\author{
         A. Jim\'enez Mu\~noz\inst{\ref{LPSC},\ref{Madrid}}
	\and J. F. Mac\'ias-P\'erez\inst{\ref{LPSC}} 
	\and G. Yepes\inst{\ref{Madrid}}
    \and M. De Petris\inst{\ref{Roma}}
    \and A. Ferragamo\inst{\ref{Roma}}
    \and W. Cui\inst{\ref{Madrid}}
    \and J.S. G\'omez\inst{\ref{Madrid}}
}
\institute{
	Univ. Grenoble Alpes, CNRS, Grenoble INP, LPSC-IN2P3, 53, avenue des Martyrs, 38000 Grenoble, France
	\label{LPSC}
	\and
	Departamento de F\'isica Te\'orica and CIAFF, M\'odulo 8, Facultad de Ciencias, Universidad Aut\'onoma de Madrid, 28049 Madrid, Spain
	\label{Madrid}
	\and
	I.N.A.F. - Osservatorio Astronomico di Roma, Via Frascati 33, 00040 Monteporzio Catone, Roma, Italy
	\label{Roma}
}	
\abstract{
Cluster number counts at visible and IR wavelengths will be a key cosmological probe in the next decade thanks to the Euclid satellite mission. For this purpose, cluster detection algorithm performance, which at these wavelengths are sensitive to the spatial distribution of the cluster galaxy members and their luminosity function, need to be accurately characterized. Using \TTH\ hydrodynamical and dark matter only simulations we study a complete sample of massive clusters beyond 7 (5) $\times$ 10$^{14}$ M$_{\odot}$ at redshift 0 (1) on a $(1.48 \ \mathrm{Gpc})^3$  volume. We find that the mass resolution of the current hydrodynamical simulations (1.5 $\times$ 10$^9$ M$_{\odot}$) is not enough to characterize the luminosity function of the sample in the perspective of Euclid data. Nevertheless, these simulations are still useful to characterize the spatial distribution of the cluster substructures assuming a common relative mass threshold for the different flavours and resolutions. By comparing with the dark matter only version of these simulations, we demonstrate that baryonic physics preserves significantly low mass subhalos (galaxies) as have also been  observed in previous studies with less statistics. Furthermore, by comparing the hydro simulations with higher resolution dark matter only simulations of the same objects and taking the same  limit in subhalo mass we find significantly more cuspy galaxy density profiles towards the center of the clusters, where the low mass substructures would tend to concentrate.  We conclude that using dark matter only simulation may lead to some biases on the spatial distribution and density of galaxy cluster members.
Based on the preliminary  analysis of few high resolution hydro simulations we conclude that a mass resolution of 1.8 $\times$ 10$^8$ h$^{-1}$ M$_{\odot}$ will be needed  for \TTH\ simulations to approach the expected magnitude limits for the Euclid survey. These simulations are currently under way. 
} 

\maketitle

\section{Introduction}
\label{sec:intro}
The abundance of clusters of galaxies \citep{press1974formation} constitutes a major cosmological probe \citep{2011ARA&A..49..409A}  for
the next generation of large-scale structure surveys like the one expected from the \Euclid satellite \citep{2011arXiv1110.3193L}. The number of clusters per unit of mass and redshift is driven by cosmological parameters as the dark matter and dark energy densities, $\Omega_{m}$ and $\Omega_{\Lambda}$ as well as the rms of the linear matter fluctuations at 8 Mpc scales, $\sigma_{8}$, via the halo mass function, which can be computed from numerical simulations \citep[e.g.][]{tinker2008toward}. A  large number of studies have been performed with multi-wavelengths observations of galaxy clusters confirming their potential as cosmological probes in  X-ray \citep{2022A&A...661A...2L,2018A&A...620A..10P,2018A&A...620A...5A,2017A&A...608A..65B}, in the optical \citep{2020PhRvD.102b3509A,2018ApJS..235...33D,2019MNRAS.485..498M}, and via the thermal Sunyaev-Zeldovich effect \citep[tSZ,][]{2021ApJS..253....3H,2019ApJ...878...55B,2016ApJ...832...95D,2015ApJS..216...27B,2016A&A...594A..24P,2014A&A...571A..20P}. Nevertheless, to date cluster cosmological constraints seem to be dominated  by systematic effects related to the observational characterization of their mass \citep[see summary in][]{pratt2019} and redshift \citep[e.g.][and references therin]{benitez2000bayesian}.  \\

One key aspect for cluster cosmology is the determination of the so called Selection Function (SF) of the survey, which gives  the probability of finding a cluster at a certain mass and redshift \citep[see for example][and references therin]{adam2019euclid}.
The SF is an intrinsic characteristic of the cluster survey and depends on the cluster finder algorithm, the observational and quality cuts, the characteristics of the survey, and the chosen observables to estimate the cluster mass (X-ray emission, tSZ effect, richness, lensing, velocity dispersion) and redshift (photometric or spectroscopic). The SF can be estimated from simulations either from full mock galaxy catalogues including clusters and field galaxies, or from individual simulated clusters that are injected in the observed galaxy survey \cite[e.g.][]{2016MNRAS.459.1764S,2019MNRAS.485..498M,adam2019euclid,2014ApJ...785..104R}. In both cases realistic physical properties of the clusters are needed over large ranges in mass and redshift and for a large variety of physical conditions.  \\

For cosmological studies  large sky volumes are needed. However, it is very difficult to produce  full hydrodynamic simulations with sufficient resolution for such large volumes. A possible solution to this problem is the `zoom-in' technique, as adopted by \TTH\ collaboration \citep{cui2018three}. For this project, a large cosmological volume is simulated by N-body dark-matter-only simulation, and only in the regions where a galaxy cluster is found, full-physics simulations are performed. 
For having enough statistics it is necessary to run a large number  of independent simulations (324 regions for the \TTH). From these simulations clusters properties can be derived and used to complement large volume dark matter only simulations \citep{2023MNRAS.518..111D}. Alternatively, the individual cluster simulations and derived cluster properties can also be used to create synthetic clusters, which can be injected in real datasets. \\

Within the perspective of cluster cosmology with the next generation of optical and infrared large scale structure surveys, and in particular of the Euclid satellite mission, we concentrate in this paper in the study of cluster member galaxies in the \TTH.  We will consider their luminosity function and spatial distribution, which are expected to drive cluster detection algorithms \cite{adam2019euclid}. In Section~\ref{sec:data} we describe the \TTH\ data used.  Sect.~\ref{sec_ch4:resoanalysis} discusses resolution effects in the determination of the luminosity function. In Section~\ref{sec_ch4:galdens} we present results in the spatial distribution of cluster galaxy members. We finally conclude in Section~\ref{sec:conclu}.

\section{\TTH\ data}
\label{sec:data}

\subsection{\TTH\ simulations}

The \TTH\ simulations were derived from the MDPL2 MultiDark Simulations \citep{2016MNRAS.457.4340K}. The latter consists on a 1 h$^{-1}$ Gpc cube containing 3840$^3$ dark matter (DM) particles with a mass of 1.5 $\times$ 10$^9$ h$^{-1}$M$_{\odot}$ each. Once the dark-matter-only simulations are performed, a cluster finder algorithm is ran. In this case the ROCKSTAR halo finder \citep{behroozi2012rockstar}, which will look for dark matter halos. A total of 324 spherical regions were extracted from the halo finder results, selecting as center for these regions the position of the most massive halo at redshift $z = 0$. The radius of each spherical region is 15 h$^{-1}$ Mpc that is much larger than the virial radius of the central cluster, which is the radius that encloses the mass that corresponds to approximately 98 times of the critical density of the Universe (at $z = 0$), as given by the Spherical Collapse model. The phase space initial conditions for the 324 selected regions are used to perform the `zoom' re-simulations. For the study presented in this paper, the 300th collaboration has ran these simulations in four different flavors based on the GIZMO-SIMBA \citep{2022MNRAS.514..977C} code, which can generate both dark matter only or hydrodynamical simulations. The GIZMO-SIMBA code is based on the Meshless Finite Mass (MFM) method. The MFM is a Langrangian method for hydrodynamics based on a kernel discretization of the volume coupled to a high-order matrix gradient estimator and a Riemann
solver acting over the volume ‘overlap’, first proposed by \citet{2015MNRAS.450...53H}. In the case of hydrodynamical simulations the GIZMO-SIMBA code applies the following baryonic physics models:

\begin{itemize}

    \item[\textbullet] The gas treatment consists of an homogeneous UV background  \citep{2012ApJ...746..125H} accounting for self-shielding on the fly \citep{2013MNRAS.430.2427R} and gas metal dependent cooling \citep{2017MNRAS.466.2217S}.

    \item[\textbullet] Star formation and stellar feedback are included using a stellar model by \citet{2016MNRAS.462.3265D} and galactic stellar and substellar initial mass function by \citet{chabrier2003galactic}. Decoupled two-phase winds with mass loading factor scaling with stellar mass and wind velocity limit in jet mode of 7000 km/s are also implemented.

    \item[\textbullet] Finally, GIZMO-SIMBA includes black hole seeding and growth and active galactic nuclei (AGN) feedback \citep{2019MNRAS.486.2827D}.\\
    
\end{itemize}

\subsection{Dataset}
\label{sec:dataset}
In this paper we have used four different flavours of \TTH\ simulations  depending mainly on the resolution and on the physics used to re-simulate the cluster regions:
\begin{enumerate}

    \item \LRD\: Dark matter only simulations with a dark matter particle resolution of 1.5$\times$10$^9$ h$^{-1}$M$_{\odot}$. 
    
    \item \HRD\  Dark matter only simulations at high resolution. They have twice more particles per dimension than the \LRD\ for a total of 7680$^3$ particles, and consequently eight times less mass per particle i.e., 1.8$\times$ 10$^8$ h$^{-1}$ M$_{\odot}$ each. 
    
    \item \LRH\ Full-physics hydrodynamics simulations at the resolution of the \LRD.
    
    \item \HRH\ Full-physics hydrodynamics simulations at the resolution of \HRD. For these simulations we only have 5 regions, due to the high computational cost. 
    
\end{enumerate}


Once the regions are re-simulated, they are analysed by the Amiga's Halo Finder \citep[AHF;][]{knollmann2009ahf}, producing a catalogue with the halos found within the different regions. In the case of hydrodynamical simulations, for each halo different properties are computed, such as its radius $R_{200}$\footnote{Radius at which the mean spherical density of the cluster is 200 times the critical density of the Universe at the cluster redshift}, mass $M_{200}$, density profile, galaxy luminosities for several spectral bands covering from far-UV to radio. The galaxy luminosities are computed from the identified stars of the AHF finder
using the STARDUST code \citep{1999A&A...350..381D}. The spectral
energy distribution (SED) of each galaxy is convolved with the bandpass of each photometric filter to compute the galaxy luminosity. In the case of dark matter only simulations we have the same properties except those related to baryon physics (e.g., stellar mass and luminosities). Each halo can have smaller halos gravitationally bounded to it, which we will call subhalos, with their own properties. The more massive and central halo is known as the host halo. 
A low mass threshold of $8\times 10^{14}\mathrm{h}^{-1}$ M$_{\odot}$ at $z = 0$ is imposed for the central halo. 
For each simulation flavours the final data consists of a halos and substructure catalogue per region.


\begin{figure*}[h!]
\centering
    \includegraphics[width=0.8\linewidth]{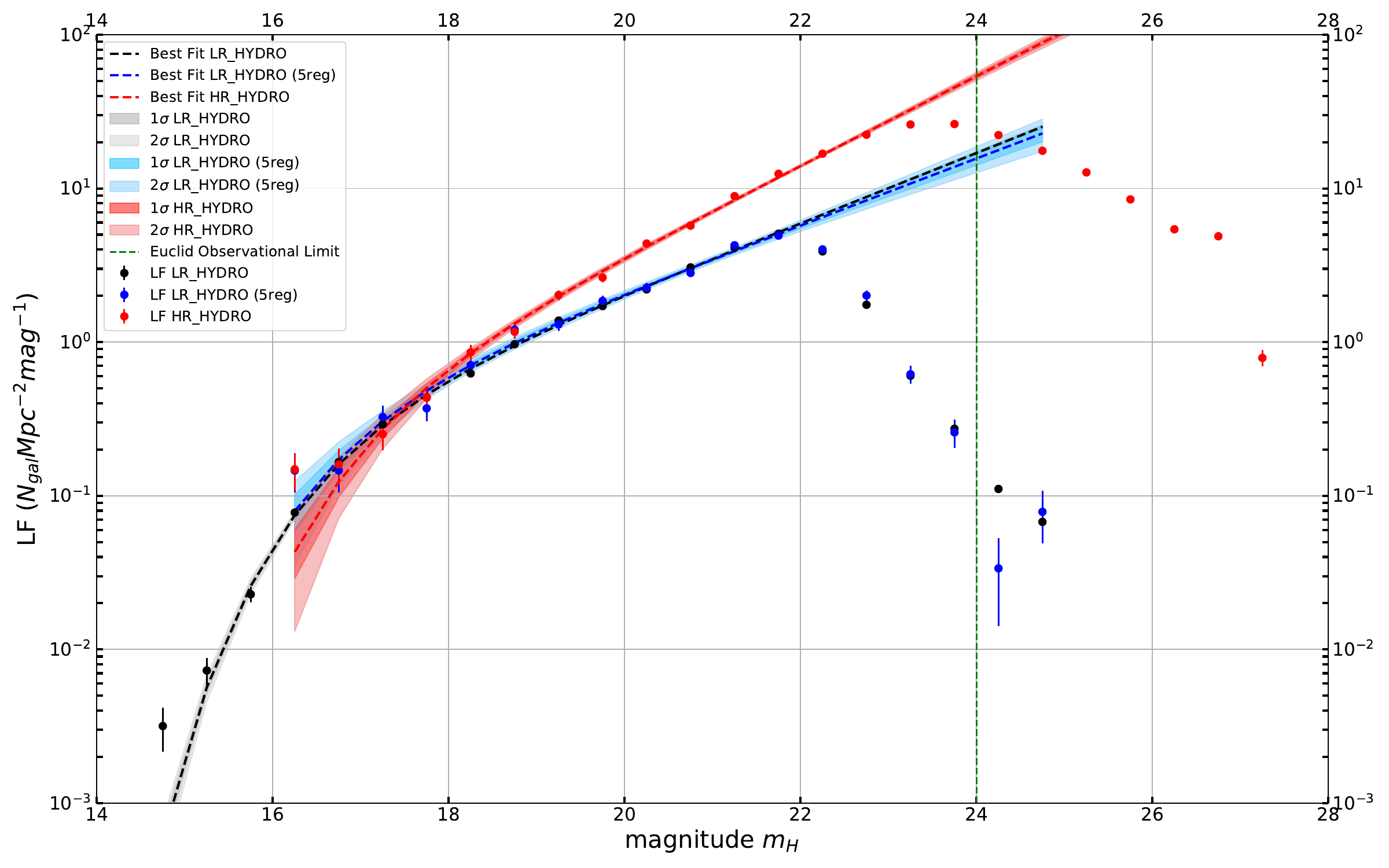}
\caption{Luminosity function in the H-band for the 324 \LRH\ regions (black), for five \HRH\ regions (red) and for the same five regions for \LRH\ (blue) at z = 1. The dots and the associated uncertainties are computed from the mean and dispersion in the bins in magnitude using all available clusters. The shaded areas are the 1$\sigma$ and 2$\sigma$ uncertainties for the best  Schechter model fit, respectively. The vertical dashed line represents the observational limit for Euclid.}
\label{fig:luminosityfunction}
\end{figure*}

\begin{figure*}[h!]
\centering
    \subfloat[z=0.0]{{\includegraphics[width=0.32\linewidth]{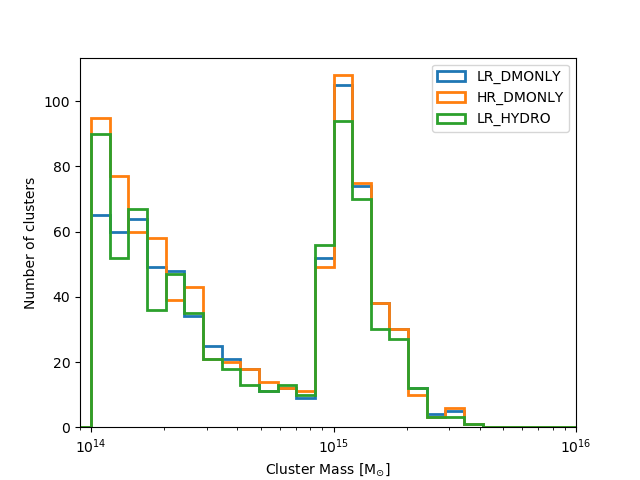}}}
    \subfloat[z=0.3]{{\includegraphics[width=0.32\linewidth]{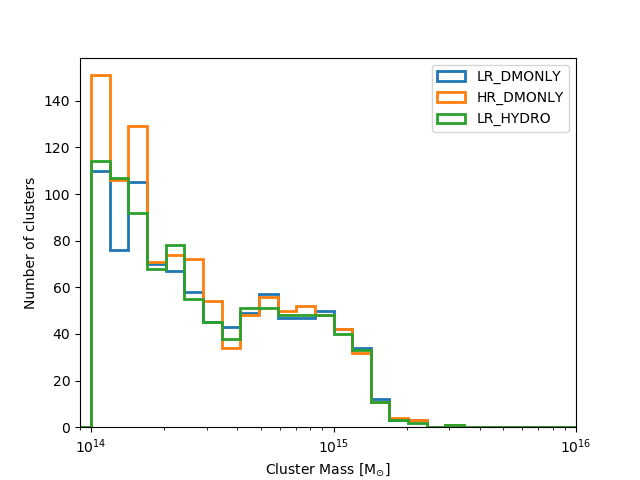}}}
    \subfloat[z=0.5]{{\includegraphics[width=0.32\linewidth]{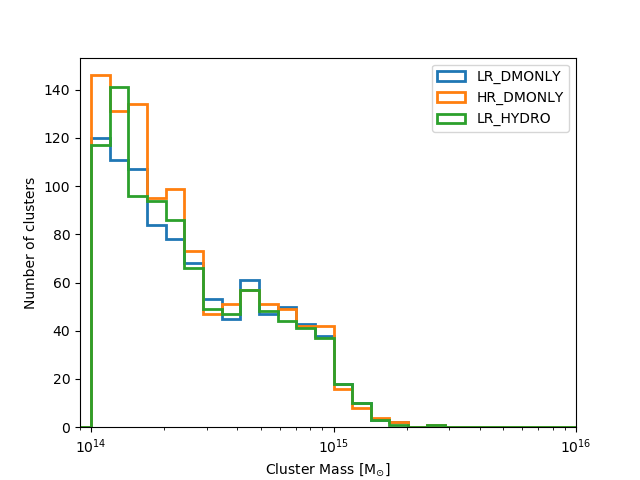}}} \hfill
       \subfloat[z=0.8]{{\includegraphics[width=0.32\linewidth]{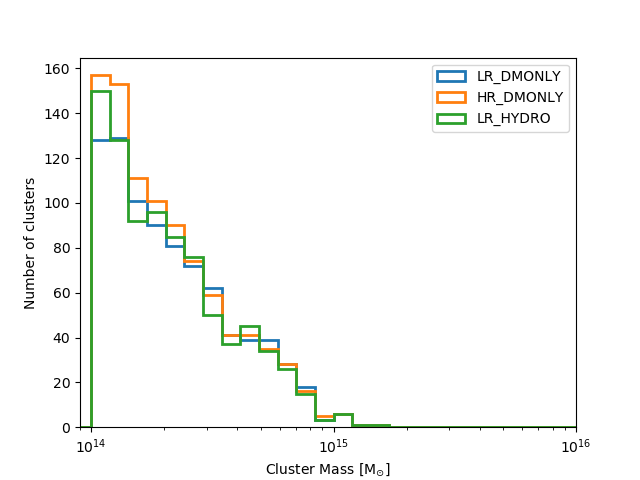}}}
    \subfloat[z=1.0]{{\includegraphics[width=0.32\linewidth]{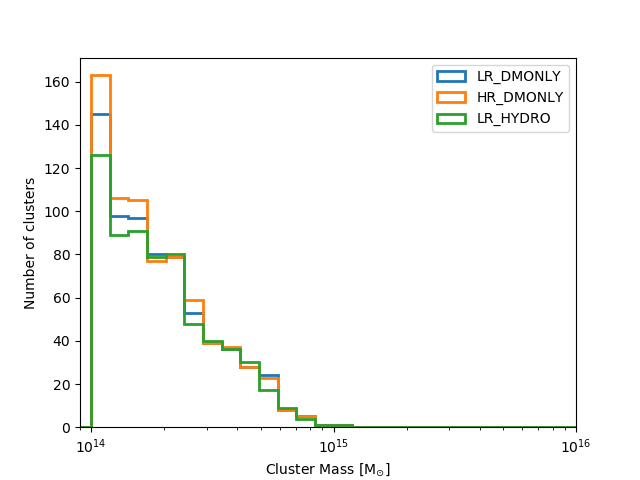}}}
     \subfloat[z=1.4]{{\includegraphics[width=0.32\linewidth]{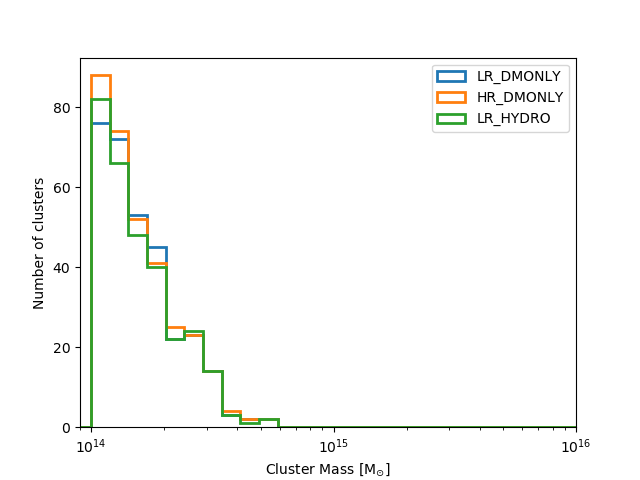}}}

\caption{Mass distribution of the selected clusters for the different snapshots in redshifts for the \LRD\ (blue), \HRD\ (orange) and \LRH\ (green) datasets described in the text. }
\label{fig:cmass_histogram}
\end{figure*}

\subsection{Identifying Galaxy Cluster Members.}
\label{subsec_ch4:dataset}



To define a galaxy, first we consider a mass threshold, for the subhalo mass, for each of four simulations. This translates into a mass resolution limit. Notice that we consider the particle mass instead of the number of particles as our threshold because for the different simulations the particle mass is different. Thus, the same number of particles does not translate into the same subhalo mass. Taking into account the particle mass and the simulation resolution we adopt the following cuts:

\begin{itemize}

    \item[\textbullet] For \LRD\, we consider 3 x 10$^{10}$ h$^{-1}$M$_\odot$ as the lower limit. This is equivalent to considering that the substructure is formed by, at least, 20 dark matter particles.
    
    \item[\textbullet] For \HRD\, considering the same mass threshold as for the low resolution ones, it leads to be at least 160 dark matter particles, because the resolution is 8 times higher. Since the resolution is significantly higher, we can vary this threshold with respect to the low resolution ones. So to check resolution effects, we choose a mass threshold of 9 x 10$^{9}$ h$^{-1}$M$_\odot$, corresponding to 50 dark matter particles.

    \item[\textbullet] \LRH\ shares resolution with the \LRD\ case. However the particle mass is different because in this case we have dark matter particles and gas particles. This means that the same mass threshold between \LRD\ and \LRH\ does not translate into the same number of particles. A hydrodynamical simulated structure can have the same mass as a dark-matter-only one, but without dark matter particles. We have to ensure that a galaxy has dark matter particles. For this reason we apply, on top of the mass threshold, a number of particles threshold. Thus we consider a mass threshold of 3 x 10$^{10}$ h$^{-1}$M$_\odot$ (as for the \LRD\ case) and a minimum number of particles threshold of 10. 
    In this way, we assure the presence of dark matter components in the baryonic structures, and also a minimum mass for having enough resolution. 

    \item[\textbullet] \HRH\ shares resolution with the \HRD\ case. So as in the latter, we choose a mass threshold of 9 x 10$^{9}$ h$^{-1}$M$_\odot$. However, following the reasoning of the \LRH\ case, instead of having a minimum of 50 dark matter particles as for \HRD\, we choose 30 dark matter particles.
    
\end{itemize}

The next step is to identify for each re-simulated region the substructures (galaxies) associated to the halos (clusters). The output of the AHF algorithm is a list of structures with their associated physical properties like for example the total, gas, and stellar mass, as well as the halo to which is gravitationally bounded. For each cluster we define as galaxy members those structures bounded to it.
After this process, our dataset is formed by a list of clusters with galaxies bounded to them. In the case of hydrodynamical simulations, we also need to check the mass ratio between the dark matter and the stellar content of the galaxy. The mass of a real galaxy is mainly coming from the dark matter halo surrounding the stars and gas, so we impose that the stellar mass component is not higher than $30\%$ of the total mass of the galaxy.
We finally get rid of contaminated low resolution particles that initially were outside the region of interest in order to maintain only well resolved structures. From now in this paper, we use subhalos or substructures, interchangeably, to refer to galaxies. The same applies to halos or clusters, to refer to galaxy clusters.\\

\begin{figure*}[!h]
\centering
    \subfloat[\BMONE]{{\includegraphics[width=0.48\linewidth]{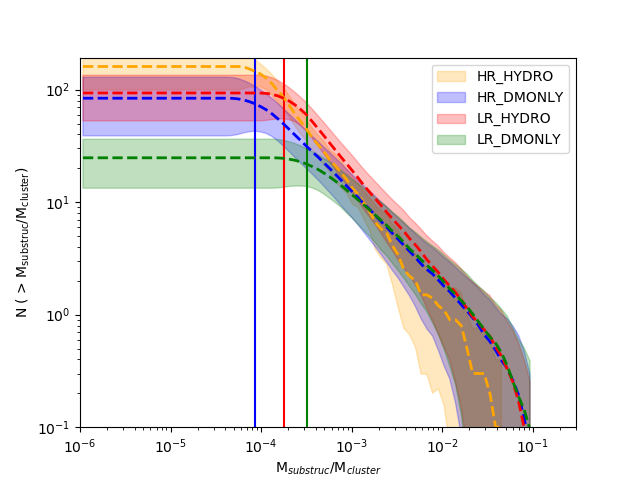}}}
    \subfloat[\BMTWO]{{\includegraphics[width=0.48\linewidth]{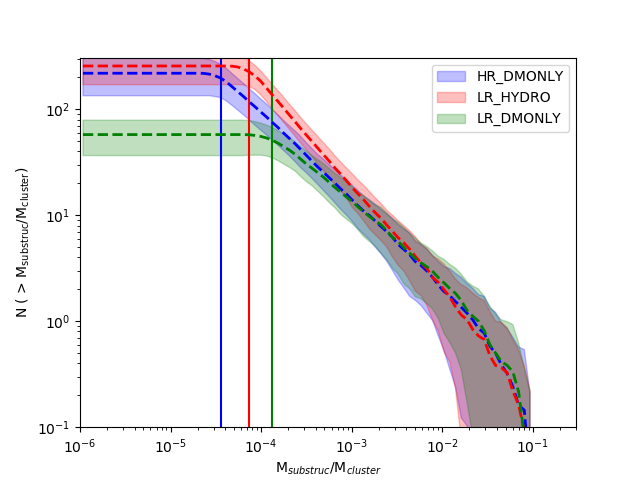}}} \hfil
       \subfloat[\BMTHREE]{{\includegraphics[width=0.48\linewidth]{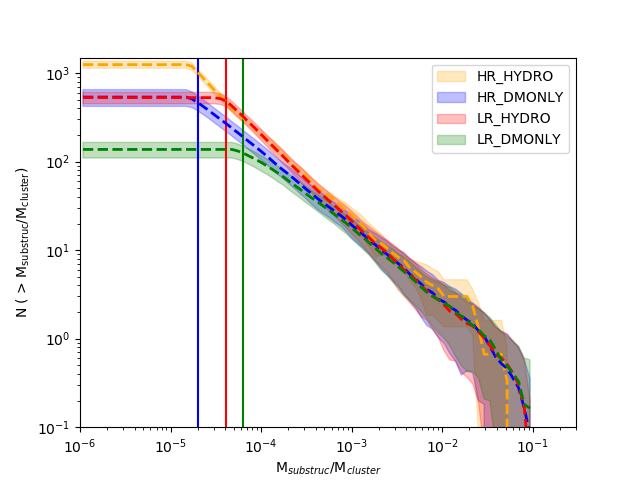}}}
    \subfloat[\BMFOUR]{{\includegraphics[width=0.48\linewidth]{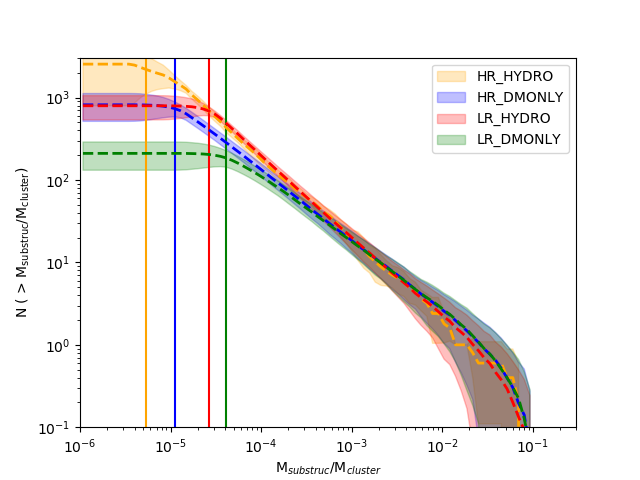}}}

\caption{Cumulative galaxy mass function for the \HRH, \HRD, \LRH\ and \LRD\ simulations at redshift 0 for the four bins in mass considered. The shaded regions correspond to the standard deviation across clusters. Resolution effects are clearly visible for the low mass region. The vertical lines represent the minimum relative mass necessary to avoid resolution effects.}
\label{fig_ch4:all_subhalo}
\end{figure*}

\begin{figure*}[h!]
\centering
    \captionsetup[subfigure]{labelformat=empty,position=top}
     \subfloat[\LRH]{
    \includegraphics[width=0.33\linewidth]{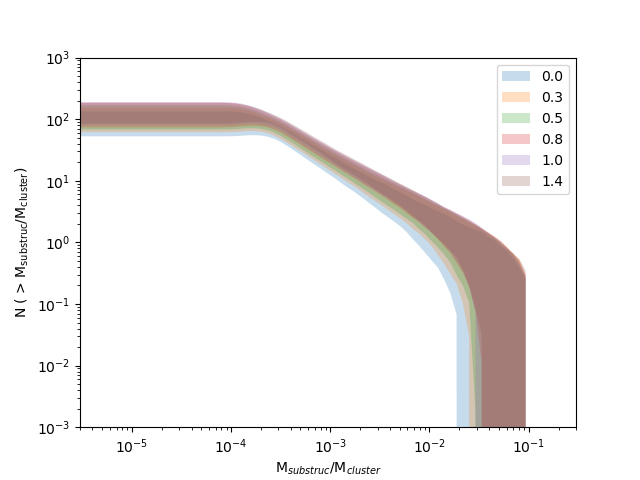}}\hfil
     \subfloat[\LRD]{\includegraphics[width=0.33\linewidth]{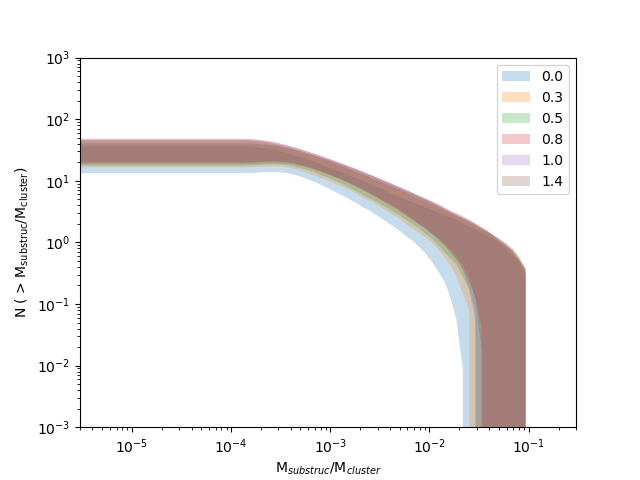}\hfil}
    \subfloat[\HRD]{\includegraphics[width=0.32\linewidth]{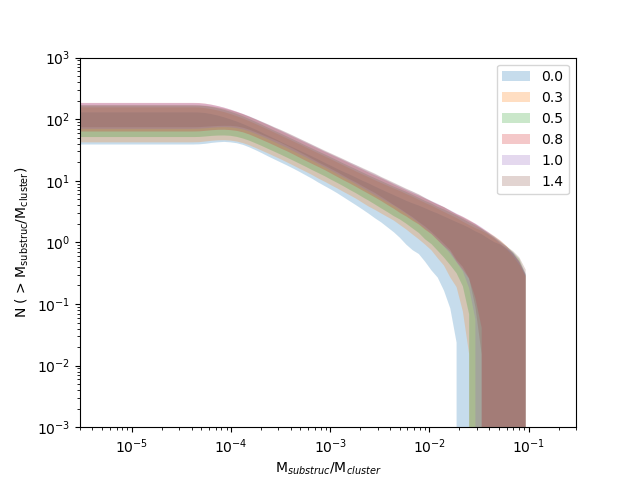} \hfil}

    \includegraphics[width=0.33\linewidth]{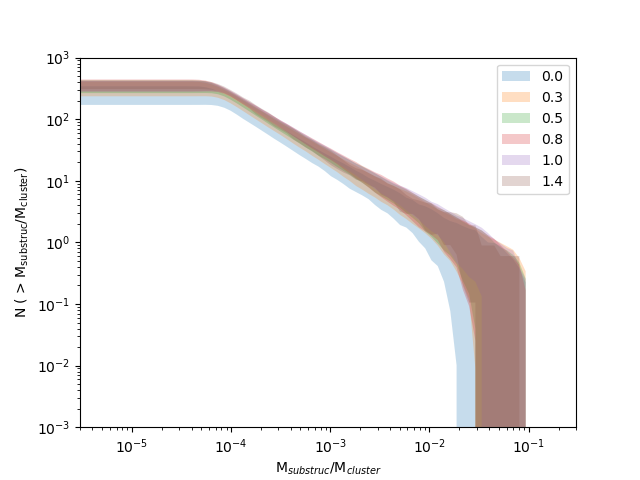}\hfil
    \includegraphics[width=0.33\linewidth]{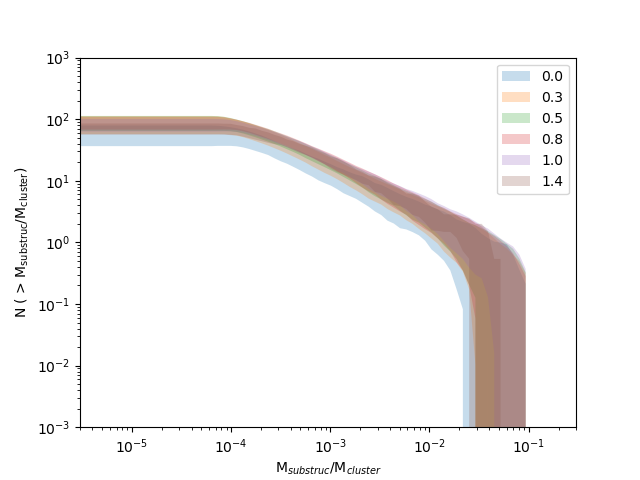}\hfil
    \includegraphics[width=0.33\linewidth]{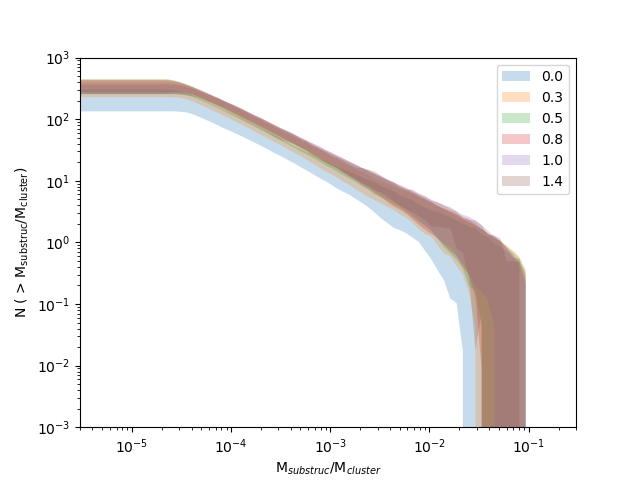}\hfil
    
     \includegraphics[width=0.33\linewidth]{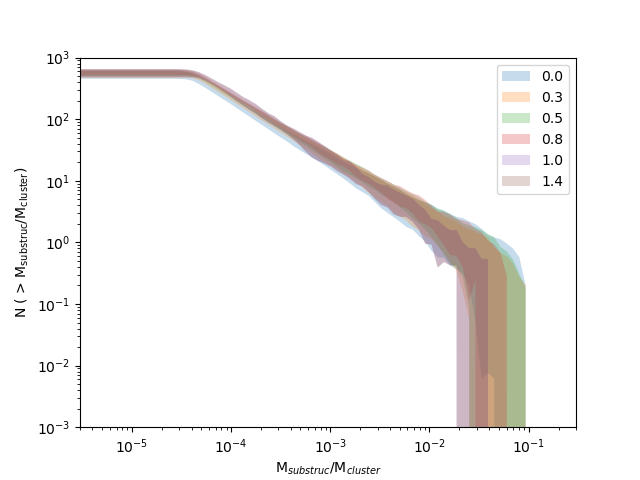}\hfil
    \includegraphics[width=0.33\linewidth]{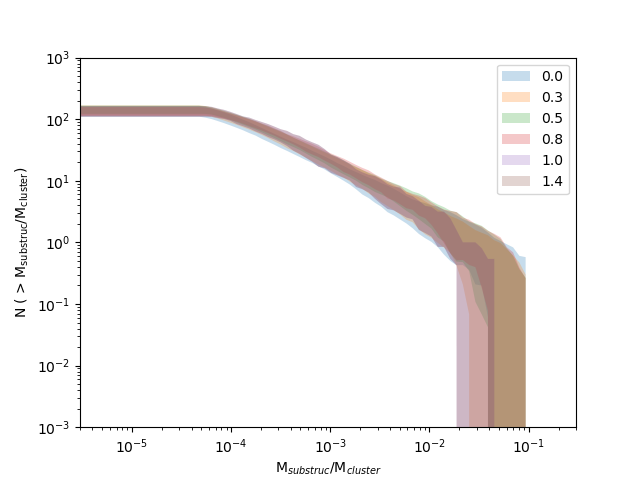}\hfil
    \includegraphics[width=0.33\linewidth]{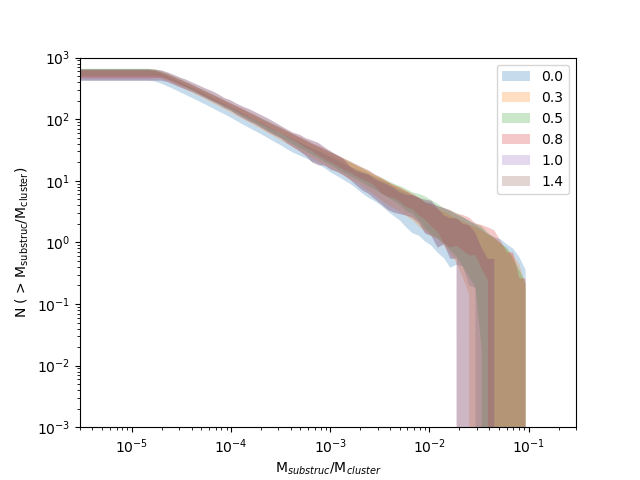}\hfil

     \includegraphics[width=0.33\linewidth]{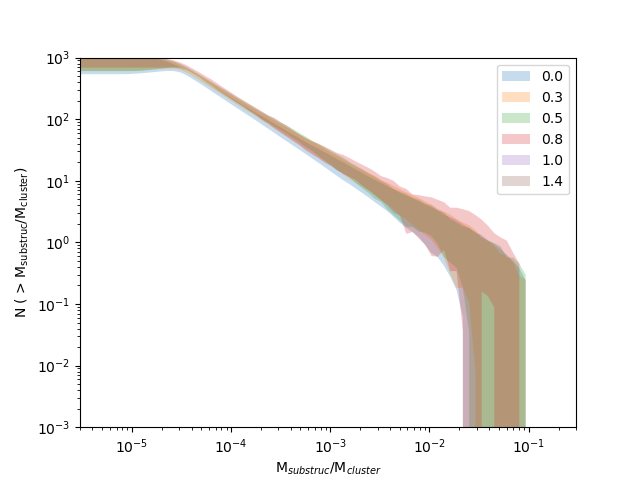}\hfil
    \includegraphics[width=0.33\linewidth]{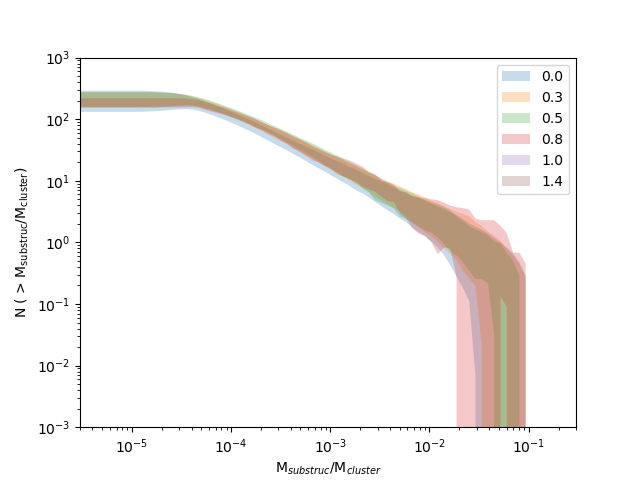}\hfil
    \includegraphics[width=0.33\linewidth]{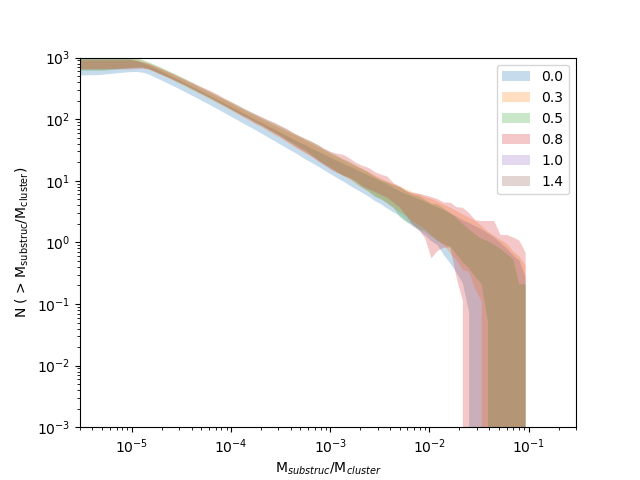}\hfil
    
 \caption{Redshift evolution of the cumulative galaxy mass function for the \LRH\ (left),  \LRD\ (center) and \HRD\ (right) simulations for the four bin mass defined in Table~\ref{table_ch4:mass_bins} (from top to bottom). The shaded regions are obtained from the mean and standard deviation across clusters as described in the text.}
\label{fig_ch4:allz_subhalo}
\end{figure*}

\begin{figure*}[h!]
\centering
    \includegraphics[width=0.38\linewidth]{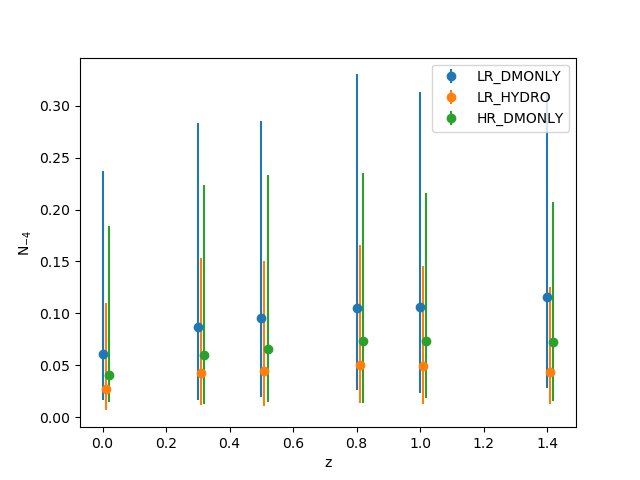}
    \includegraphics[width=0.38\linewidth]{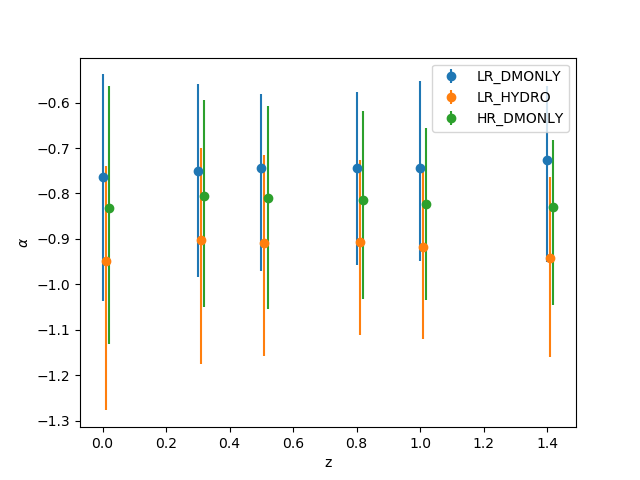}\hfil
    \includegraphics[width=0.38\linewidth]{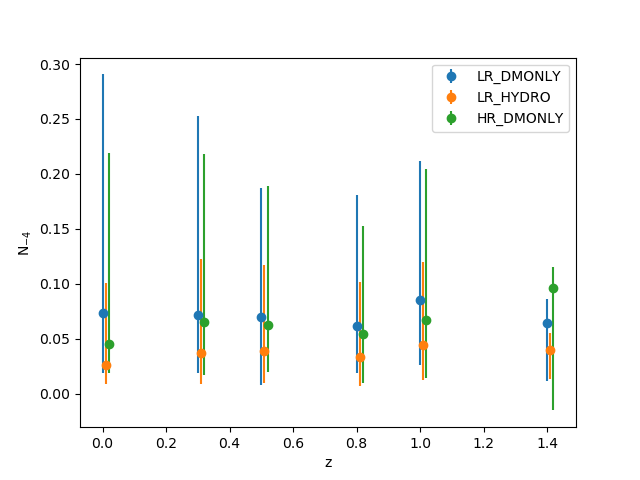}
    \includegraphics[width=0.38\linewidth]{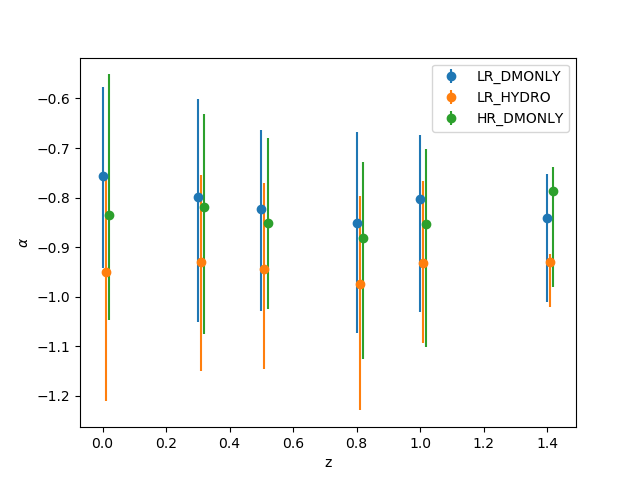}\hfil
    \includegraphics[width=0.38\linewidth]{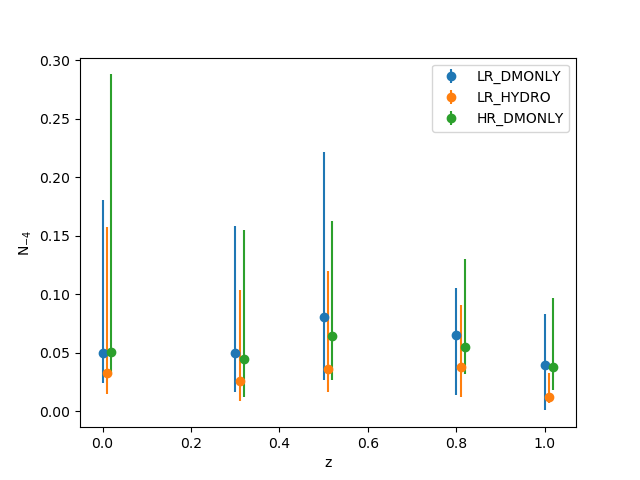}
    \includegraphics[width=0.38\linewidth]{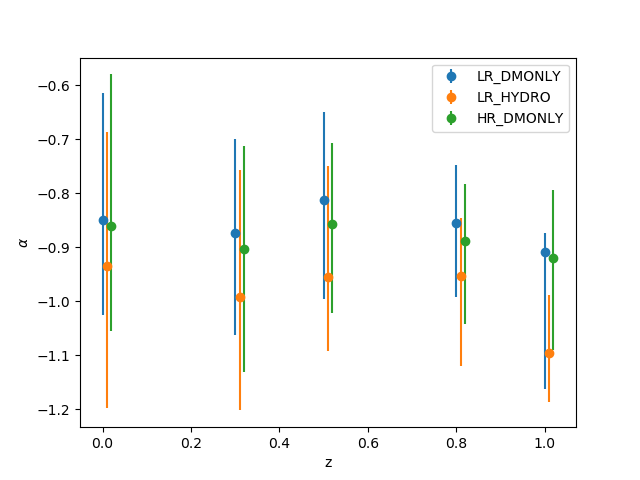}\hfil
  \includegraphics[width=0.38\linewidth]{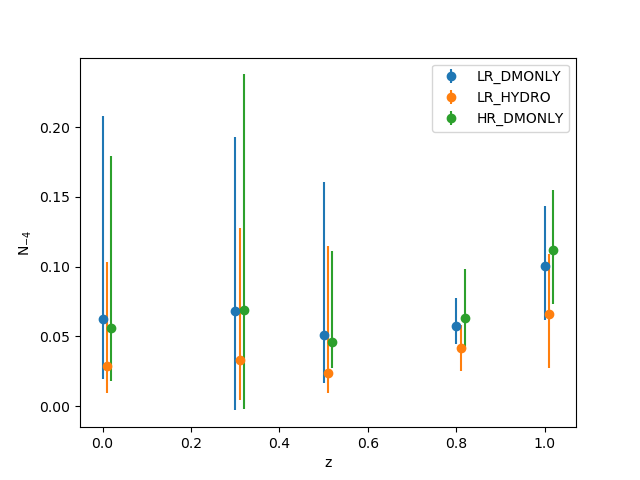}
    \includegraphics[width=0.38\linewidth]{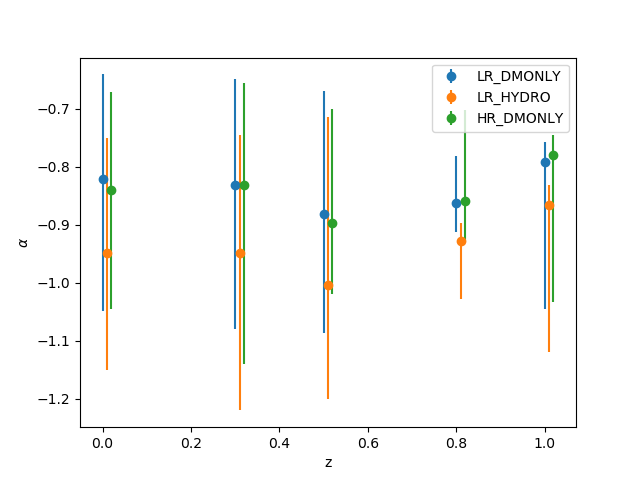} \hfil
\caption{Redshift and mass evolution of the parameters from the power-law fit of the 3D cumulative galaxy mass function. The right column represents the $\alpha$ parameter and the right column the normalization $N_{-4}$, for $z = 0$ up to $z = 1.0$. From top to bottom we present results for the different cluster mass bins of Table~\ref{table_ch4:mass_bins}.\label{fig_ch4:fit_powerlaw}}
\end{figure*}

\begin{figure*}[h!]
\centering
\includegraphics[width=0.36\linewidth,trim={0cm 0 1cm 0},clip]{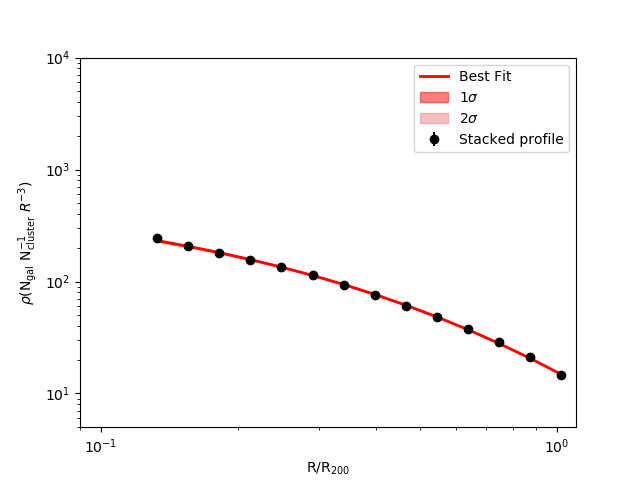}
\includegraphics[width=0.36\linewidth,trim={0cm 0 1cm 0},clip]{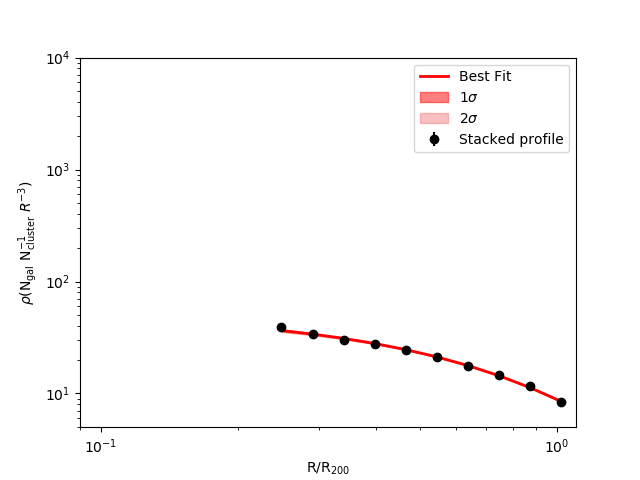} \hfil
\includegraphics[width=0.36\linewidth,trim={0cm 0 1cm 0},clip]{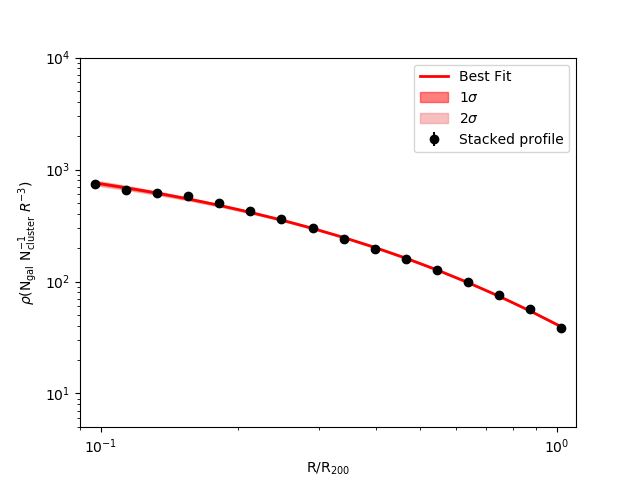}
\includegraphics[width=0.36\linewidth,trim={0cm 0 1cm 0},clip]{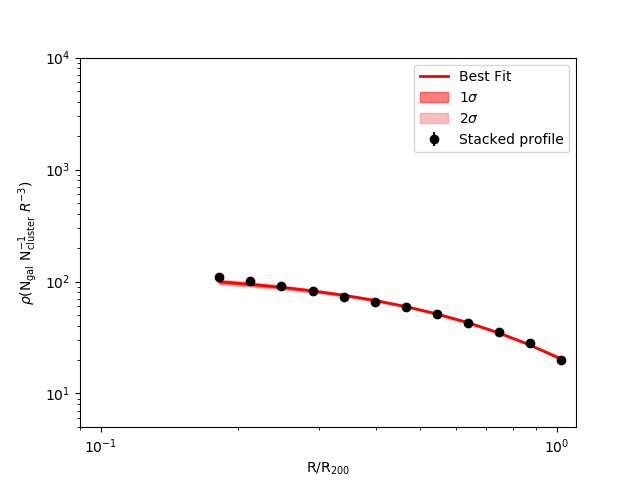} \hfil
\includegraphics[width=0.36\linewidth,trim={0cm 0 1cm 0},clip]{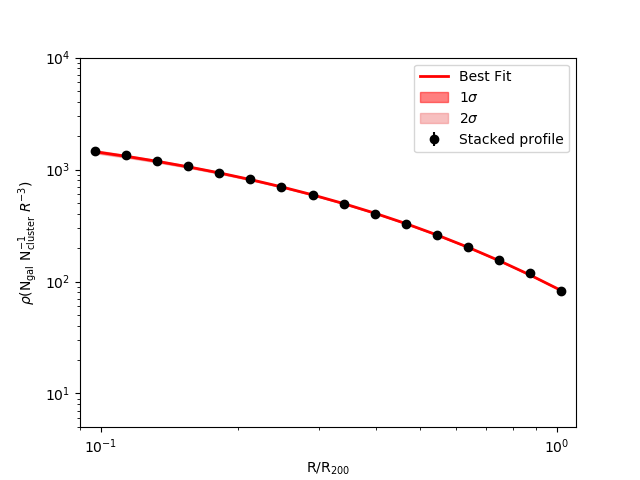}
\includegraphics[width=0.36\linewidth,trim={0cm 0 1cm 0},clip]{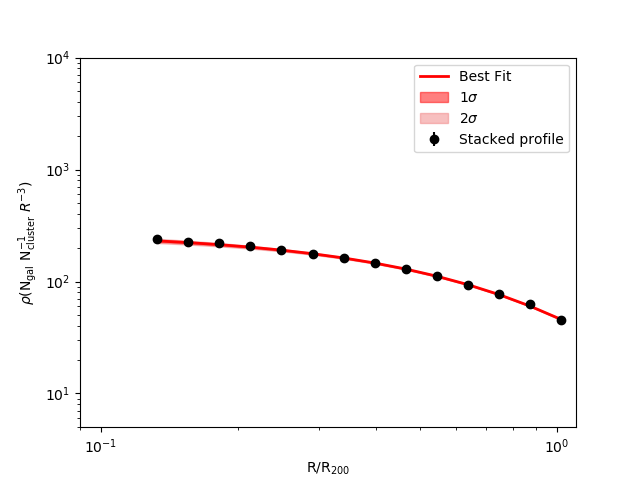} \hfil
\includegraphics[width=0.36\linewidth,trim={0cm 0 1cm 0},clip]{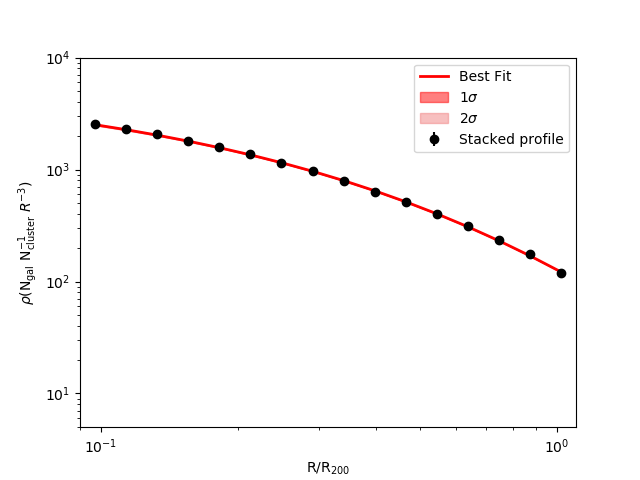}
\includegraphics[width=0.36\linewidth,trim={0cm 0 1cm 0},clip]{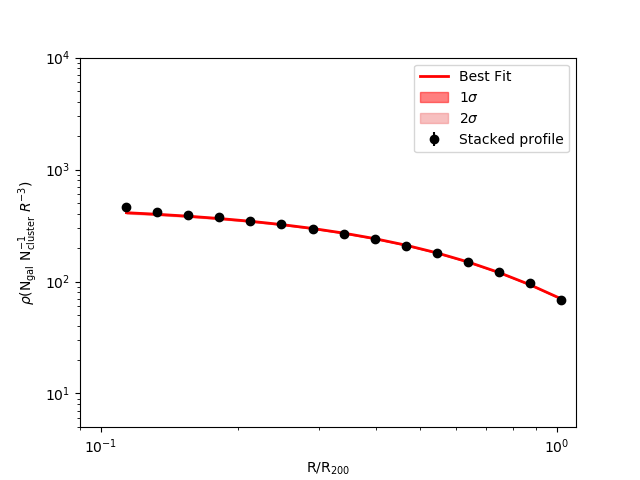} \hfil
\caption{3D cumulative galaxy density radial profile for the \LRH\ (left column) and \HRD\ (right column) simulations at redshift zero. From top to bottom we show results for the four bin in mass in Table~\ref{table_ch4:mass_bins}. The black dots and error bars represent the stacked profile in the bin and uncertainties. The red lines are the best-fit Einasto models to the stacked profiles following a MCMC approach as described in text. The dark and light red shaded areas are the 1$\sigma$ and 2$\sigma$ uncertainties for the best Einasto model fit, respectively.\label{fig_ch4:HR_HYDRO_density_profile_fit}}
\end{figure*}

\section{Resolution Effects}
\label{sec_ch4:resoanalysis}

\subsection{Luminosity Function}
\label{subsec_ch4:LF}

As discussed above a key property for the detection of clusters of galaxies in large scale structure surveys is their Luminosity Function (LF), defined as the projected density of galaxies per unit of magnitude \citep[see][for discussion]{adam2019euclid}. We have therefore computed the LF for the \LRH\ simulations as shown in Figure~\ref{fig:luminosityfunction} (black dots and associated uncertainties). For simplicity we have considered in the figure only the clusters at $z=1$. To ease the comparison to the results on the Mock Euclid catalogue presented in \citet{adam2019euclid} we compute the LF for apparent magnitudes in the H-band. First of all, we observe a clear deficit of galaxies at faint magnitudes with a clear drop at magnitude 21. This is even more obvious when trying to fit a Schechter model to the data via a MCMC analysis following \citet{adam2019euclid}. The dark and light gray shaded areas in the figure correspond to the 1$\sigma$ and 2$\sigma$ uncertainties over the best-fit Schechter model, respectively. \\


The observed drop in the number of faint galaxies is most probably due to a lack of resolution in the \LRH\ simulations. To test this hypothesis, the \TTH\ collaboration has produced five \HRH\ simulations as presented in Sect.~\ref{sec:data}. We present in red dots in the Figure~\ref{fig:luminosityfunction} the LF for several clusters in the \HRH\ simulations. We compute the LF as for the \LRH\ simulations and also fit it to a Schechter model. We observe that the LF reaches a maximum magnitude of 24 which corresponds to the observational limit expected for Euclid (green vertical line in Figure~\ref{fig:luminosityfunction}). We also observe larger uncertainties as only five regions are considered but no drop in the number of galaxies. Furthermore the Schechter model seems to be a good fit to the data as expected. We observe fainter galaxies for the \LRH\ simulations most probably due to the lack of statistics in the \HRH\ simulations. To check this, we present in blue dots in Figure~\ref{fig:luminosityfunction} the LF for the \LRH\ simulations for the same five regions as the \HRH\ simulations. Indeed, we observe that the fainter part of the LF agrees between both resolutions. From these results we conclude that the resolution of the \LRH\ simulations is not enough for studying the properties of the LF of clusters for the next generation of large scale structure surveys as Euclid. Producing \HRH\ is computationally very expensive and so for this paper we only dispose of five regions. In the following we will use them for qualitative comparison and limit our quantitative analysis to the \LRH, \LRD\ and \HRD\ simulations.

\subsection{Subhalo Mass Function}
\label{subsed:massfunction}

The subhalo mass function gives the number of halo member galaxies (subhalos) of a given mass, relative to the mass of the halo. The final mass resolution of the simulation will imprint in the subhalo mass function as a deficit of low mass subhalos. \\


\begin{center}
\begin{table}[h!]
\begin{tabular}{|c|c|} 
 \hline
  Mass Bin & Cluster Mass [10$^{14}$ h$^{-1}$M$_\odot$] \\
 \hline\hline
 \BMONE\ & 1.0  $\leq$ $M_{200}$ < 3.5 \\
 \hline
 \BMTWO\ & 3.5 $\leq$ $M_{200}$ < 7.0 \\
 \hline
 \BMTHREE\ & 7.0 $\leq$ $M_{200}$ < 10 \\
 \hline
 \BMFOUR\ &  10 $\leq$ $M_{200}$ $\leq$  100 \\
 \hline
\end{tabular}
\caption{Definition of the cluster mass bins used for the analyses presented in the main text. We give the interval in mass considered in units of 10$^{14}$h$^{-1}$ M$_\odot$. }
\label{table_ch4:mass_bins}
\end{table}
\end{center}

To compute the subhalo mass function we first divide our sample in four bins in mass as described by Table~\ref{table_ch4:mass_bins}. These were defined at redshift zero so that we can have bins as narrow as possible in mass (to ensure equivalent properties for the clusters in the bin) while preserving sufficient statistics per bin. We show in Fig.~\ref{fig:cmass_histogram} the mass distribution of the selected clusters for the 6 redshift snapshots considered in this paper. The mass histograms for the \LRH, \LRD\ and \HRD\ simulations are displayed in green, blue and orange, respectively. On the one hand, we find that the mass distribution at each redshift is very similar for the three simulation flavors. On the other hand, we observe that at high redshift the number of massive clusters decreases significantly as one would expect. This will increase the number of clusters for the lowest mass bins. \\

The resolution effect in the subhalo (galaxy) mass function can be clearly observed in Figure~\ref{fig_ch4:all_subhalo}. For the four bin in mass and at redshift zero we present the cumulative galaxy mass function for the four simulation flavors. We use all the available regions for each of the simulation flavors. The shaded regions displayed in the figure are obtained for each flavor from the mean and dispersion of the galaxy mass function of individual clusters in the mass bin. We plot the total number of substructures (galaxies or subhalos) per halo (cluster) with relative mass, $M_{\rm{substruct}}/M_{\rm{halo}}$, large than a certain threshold.
For the second mass bin the \HRH\ simulations has no cluster because of the low statistics (only five regions available). \\

We observe, in terms of resolution, for very low mass subhalos we find significant differences, with high resolution simulations showing more subhalos at low mass than their respective low resolution simulations. However, the subhalo mass function for the four simulation flavours are very similar at intermediate masses and at high mass they converge into very similar distribution within the variance. We stress that the \HRH\ simulations are shown only for qualitative comparison because of the lack of statistics. These two effects can be explained by the fact that low mass substructures can not be formed unless sufficient resolution is attained.

We also observe some physical effects. For the \LRH\ simulation we can see for all mass bins more galaxies than for the \LRD\ ones. This is probably due to baryonic physics, which diminishes the ripping out of the objects because of cooling down processes of the gas, and so permits keeping more less massive galaxies. The increase of resolution in the \HRD\ simulations tends to increase also the number of low mass substructures (galaxies). However, we observe how the total number of substructures is the same between the \HRD\ simulations and the \LRH\ ones.

\subsection{Redshift evolution of the subhalo mass function}
\label{subsec_ch4:mass_function_variation}

\begin{table*}
    \centering
    \begin{tabular}{|c|c|c|c|}
    \cline{2-4}
    \multicolumn{1}{ c|}{}      &      \multicolumn{3}{|c|}{$\left< \alpha \right>_{z}$}   \\
  \cline{1-4 }
 Mass Bin &    \LRD\ & \LRH\ & \HRD\ \\ 
 \hline
\BMONE\  & $-0.746 \pm 0.011$ & $-0.921 \pm 0.018$ & $-0.819 \pm 0.010$ \\
\BMTWO\ &  $-0.813 \pm 0.031$ & $-0.943 \pm 0.016$ & $-0.838 \pm 0.029$ \\
\BMTHREE\ & $-0.860 \pm 0.032$ & $-0.986 \pm 0.058$ & $-0.886 \pm 0.024$ \\
\BMFOUR\ &$-0.838 \pm 0.032$ & $-0.939 \pm 0.044$ & $-0.842 \pm 0.038$ \\
\hline
    \end{tabular}
    \caption{Redshift averaged and uncertainties of the slope, $\alpha$, of the subhalo mass function for the four bins in mass and for the different simulation flavors considered.}
    \label{tab:massfunctionalpha}

\end{table*}

\begin{table*}[h!]
    \centering
    \begin{tabular}{|c|c|c|c|c|c|c|}
    \cline{2-7}
          \multicolumn{1}{ c|}{} & \multicolumn{6}{|c|}{ $M_{\rm{substruc}}/M_{\rm{halo}}$} \\
             \cline{1-7}
    Mass Bin & z=0.0 & z=0.3 &  z=0.5 &  z=0.8 & z=1.0 & z=1.4 \\
    \hline
    \BMONE\ & 0.000177 & 0.000177 & 0.000177 & 0.000177 & 0.000177 & 0.000205 \\ 
    \BMTWO\ & 0.000074 & 0.000074 & 0.000074 & 0.000074 & 0.000085 & 0.000074\\
    \BMTHREE\ &0.000041 & 0.000048 & 0.000048 & 0.000048 & 0.000055 & 0.000074 \\
    \BMFOUR\ & 0.000027 & 0.000031 & 0.000031 & 0.000036 & 0.000041 & 0.000074 \\
    \hline
    \end{tabular}
    \caption{Relative galaxy to cluster mass, $M_{\rm{substruc}}/M_{\rm{halo}}$,  cut applied for comparing the \LRH, \LRD\ and \HRD\ simulations for the six different redshift snapshots.}
    \label{tab:cuthydro}
\end{table*}
We show in Fig~\ref{fig_ch4:allz_subhalo} the cumulative subhalo mass function for the \LRH\ (left column), \LRD\ (center column) and \HRD\ (right column) simulations for the six redshift snapshots considered. They have been computed as for Fig.~\ref{fig_ch4:all_subhalo} -- the shaded region being obtained from the mean and dispersion across clusters.
From top to bottom the rows correspond to the bins in mass defined in Table~\ref{table_ch4:mass_bins}. These subhalo mass functions can be compared to those presented in \citep{dolag2009substructures}. With respect to the latter we have increased in this work significantly the statistics as well as extended the cluster mass and redshift ranges considered. Furthermore, we dispose of simulations at different resolutions.


These cumulative galaxy mass functions can be well approximated by a power-law function as in \citet{dolag2009substructures},
\begin{equation}
    N_m = N_{-4}\left(\frac{ m_{sub}/M_{vir}}{M_\odot}\right)^\alpha,
    \label{eq_ch4:powerfit}
\end{equation}

\noindent where N$_{-4}$ is a normalization, $\alpha$ the slope and M$_{substruct}/$M$_{cluster}$ is the ratio between the virial mass of the substructures (galaxies) and that of their host halo. 
For obtaining the best-fit parameters we have performed a least square fits of the mean value accounting for the uncertainties computed from the dispersion across clusters. In addition, we have also performed a fit of the cumulative galaxy mass function per cluster and we compute the dispersion of the best-fit parameters across clusters for each bin in mass and for each redshift slice.\\

We present in Figure~\ref{fig_ch4:fit_powerlaw} the evolution with redshift for both, the normalization (left column) and the slope (right column) best-fit parameters for the four cluster mass bins discussed above (top to bottom). 
The color dots represent the best fit parameters for the four simulation flavors: \LRH\ (orange), \LRD\ (blue) and \HRD\ (green). The uncertainties are computed from the dispersion of the power-law fit of the 3D cumulative galaxy mass function distribution per cluster, and from the intrinsic uncertainty on the fit of the mean cumulative galaxy mass function as discussed before. We observe no evolution with redshift for both parameters as was already the case in \citet{dolag2009substructures}. However, we observe in average a slight reddening of the slope with cluster mass. This is more obvious on Table~\ref{tab:massfunctionalpha} where we present the redshift averaged slope and uncertainties for the four bins in mass.


\begin{figure*}[h!]
\centering
    \includegraphics[width=0.35\linewidth]{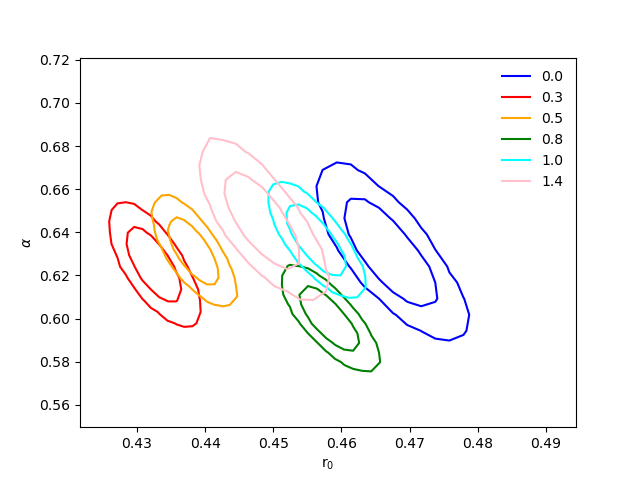}
    \includegraphics[width=0.35\linewidth]{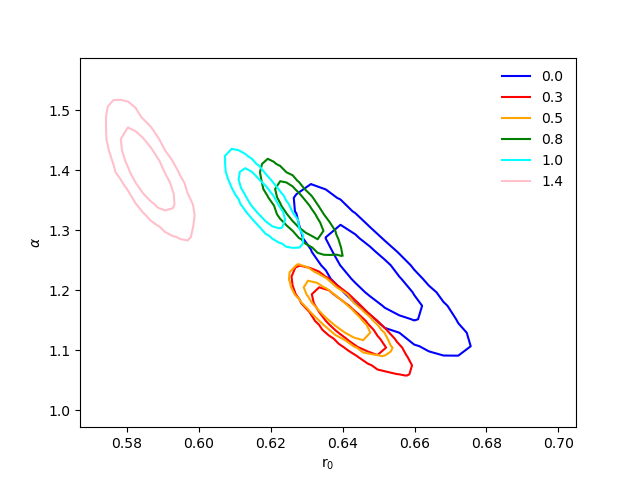}\hfil
    \includegraphics[width=0.35\linewidth]{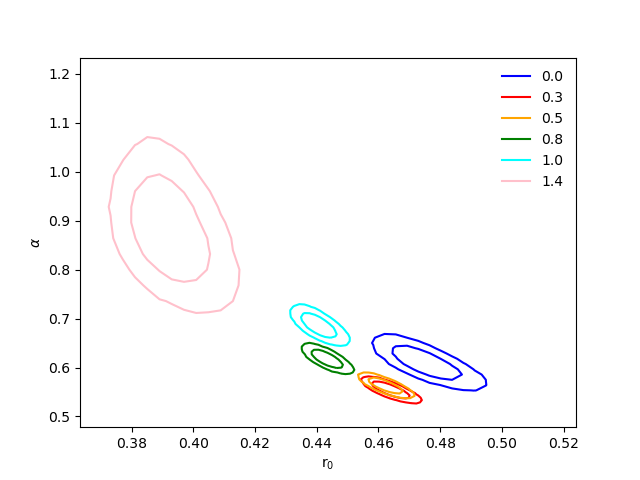}
    \includegraphics[width=0.35\linewidth]{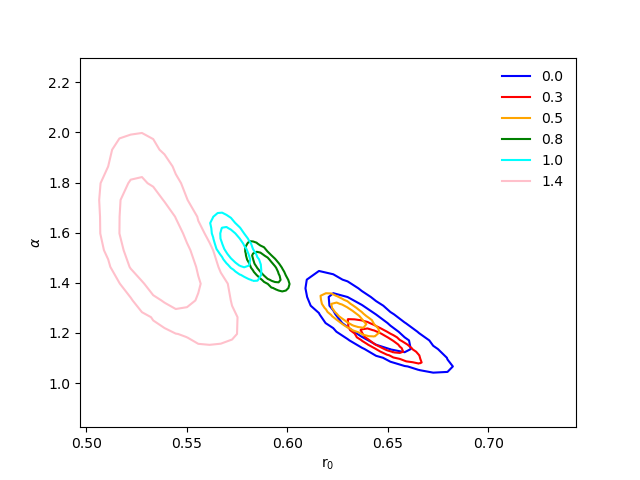}\hfil
    \includegraphics[width=0.35\linewidth]{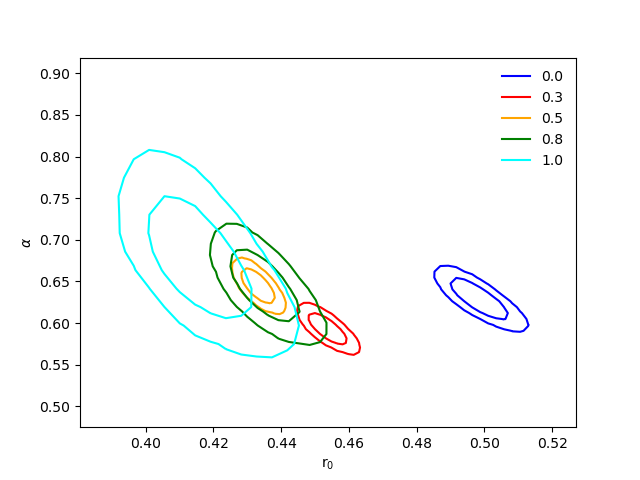} \includegraphics[width=0.35\linewidth]{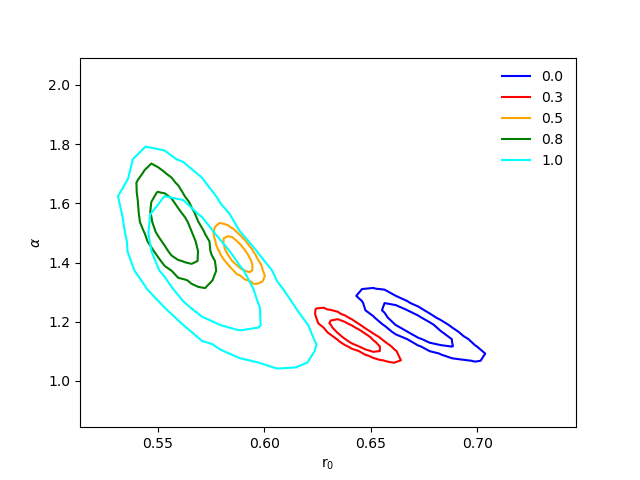}\hfil
    \includegraphics[width=0.35\linewidth]{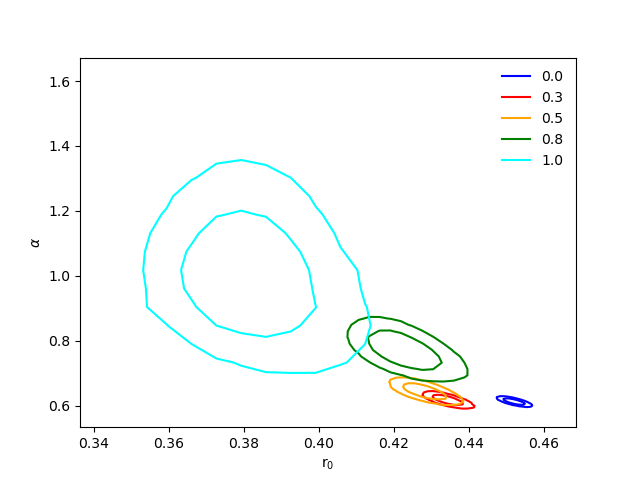}
    \includegraphics[width=0.35\linewidth]{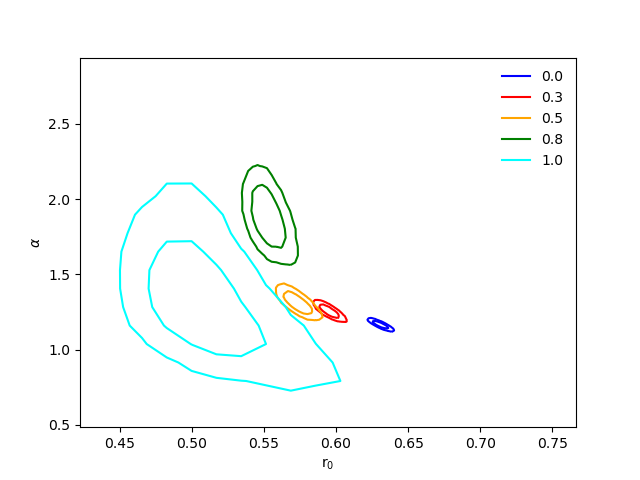}\hfil
\caption{Redshift evolution of the 2D probability distribution of the $\alpha$ and r$_0$ parameters of the Einasto model for the \LRH\ (left column) and \HRD\ (right column) simulations. From top to bottom we give results for the four bin in mass in Table~\ref{table_ch4:mass_bins}. The inner and outer contours correspond to the 68~\% and 95.4~\% C.L., respectively.
\label{fig:einasto_alphar0}}
\end{figure*}

\begin{figure*}[h!]
\centering
  \includegraphics[width=0.35\linewidth]{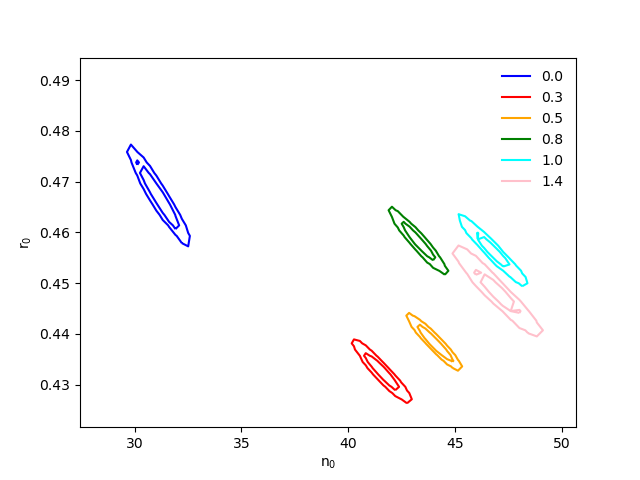}
    \includegraphics[width=0.35\linewidth]{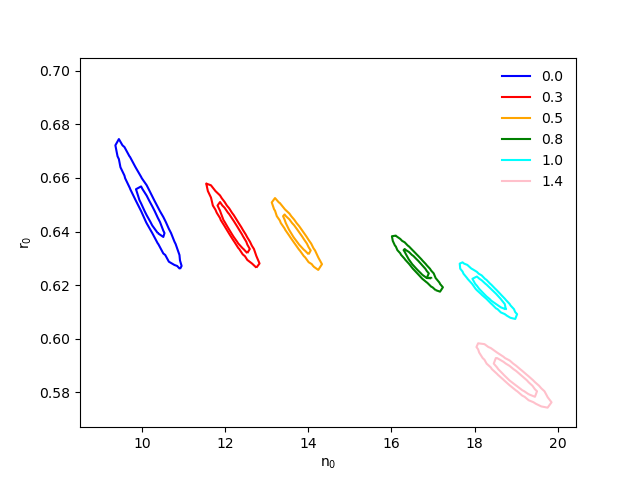}\hfil
    \includegraphics[width=0.35\linewidth]{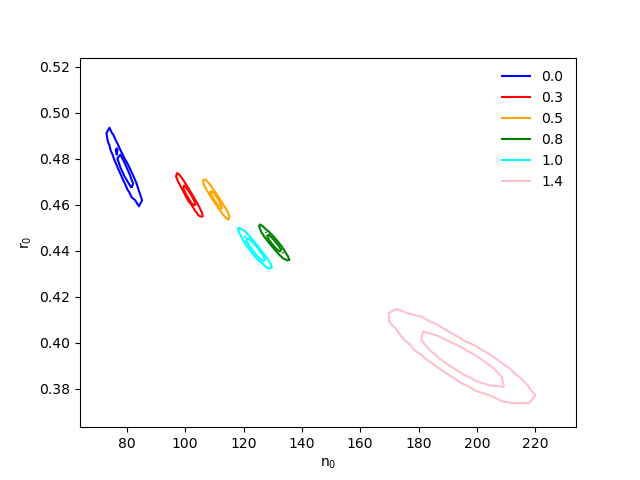}
    \includegraphics[width=0.35\linewidth]{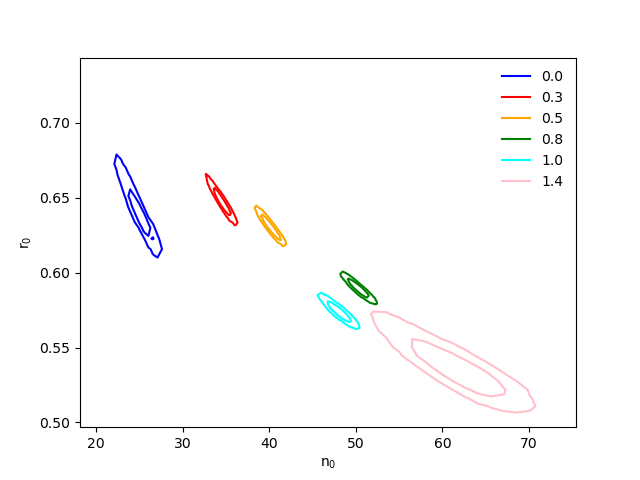}\hfil
    \includegraphics[width=0.35\linewidth]{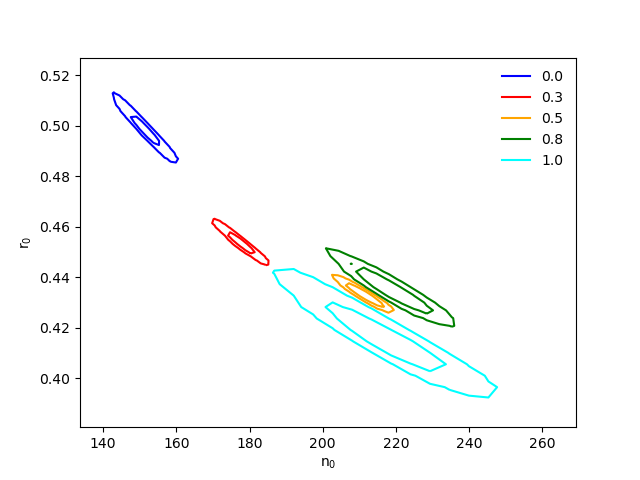}
    \includegraphics[width=0.35\linewidth]{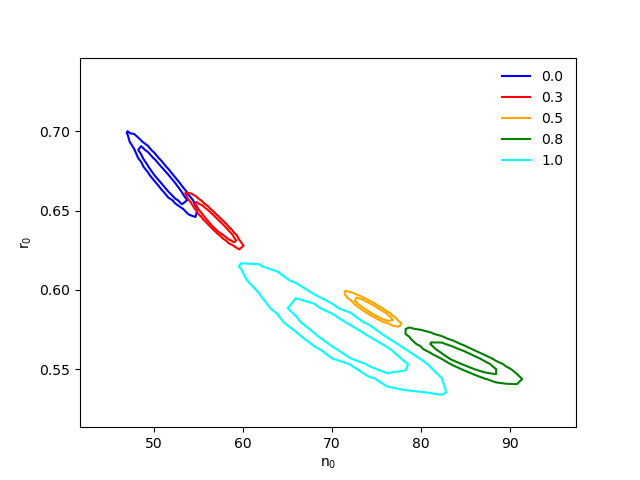}\hfil
    \includegraphics[width=0.35\linewidth]{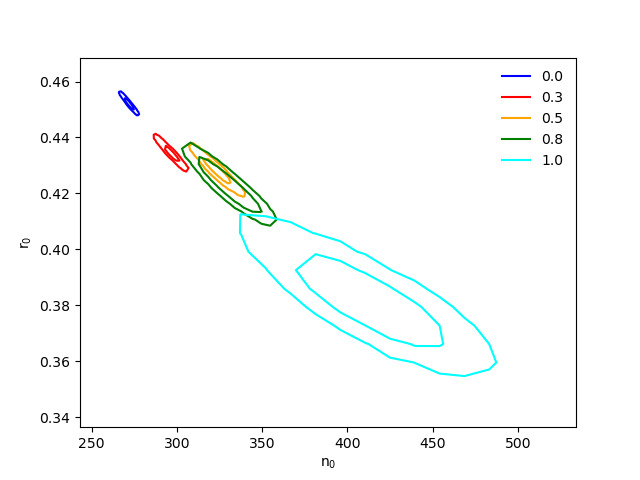}
    \includegraphics[width=0.35\linewidth]{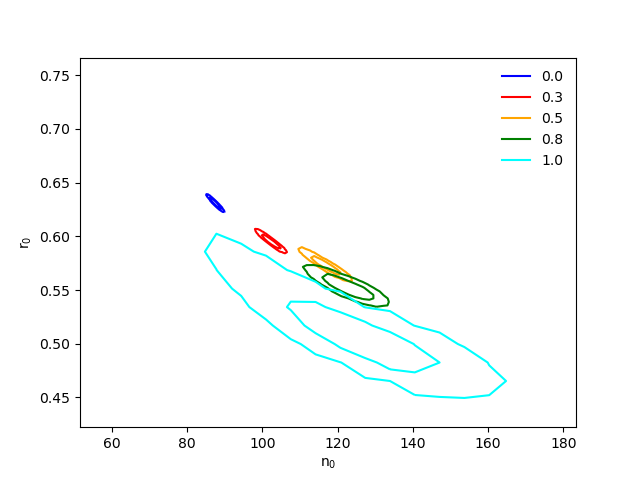}\hfil
\caption{Redshift evolution of the 2D probability distribution of the r$_0$ and n$_0$ parameters of the Einasto model for the \LRH\ (left column) and \HRD\ (right column) simulations. From top to bottom we give results for the four bin in mass in Table~\ref{table_ch4:mass_bins}. The inner and outer contours correspond to the 68~\% and 95.4~\% C.L., respectively. \label{fig:einasto_r0n0}}
\end{figure*}

\begin{figure*}[h!]
\centering 
    \includegraphics[width=0.31\linewidth]{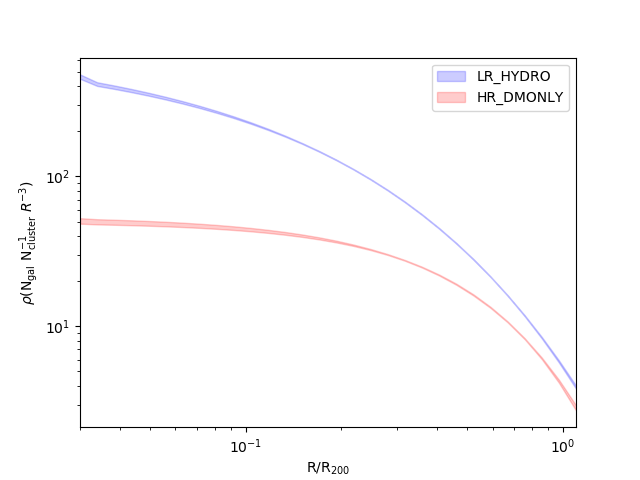}
    \includegraphics[width=0.68\linewidth]{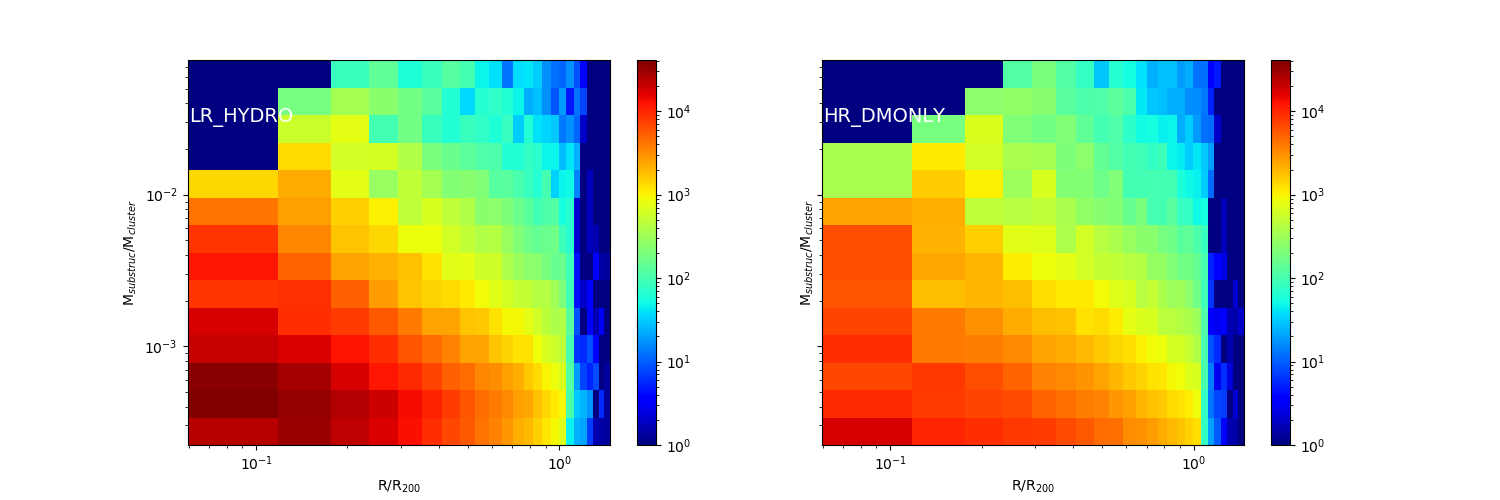}\hfil
    \includegraphics[width=0.31\linewidth]{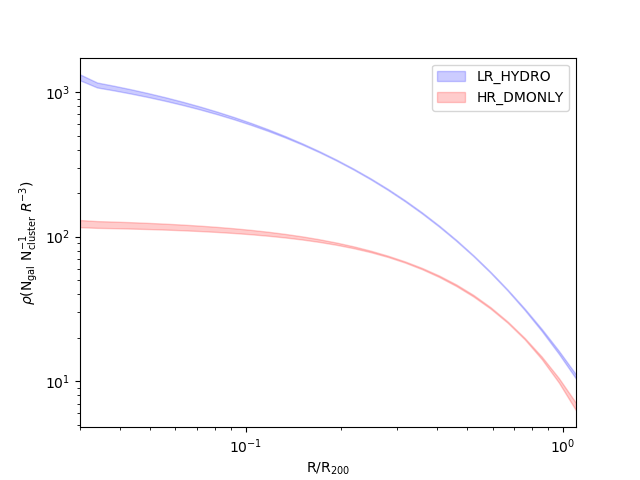}
    \includegraphics[width=0.68\linewidth]{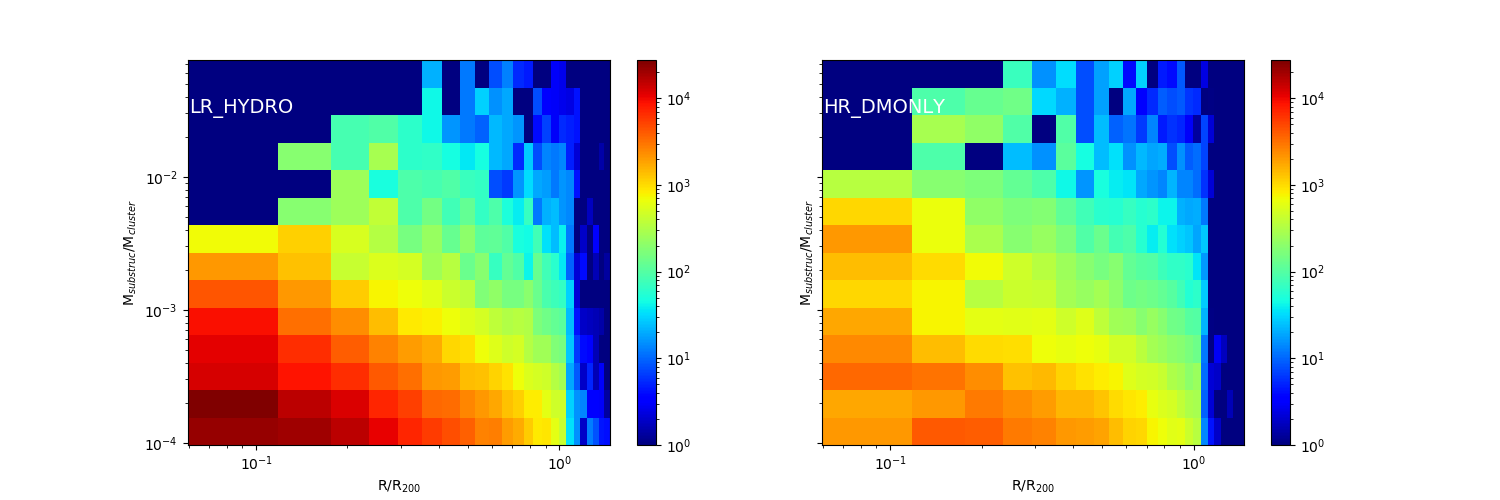}\hfil
    \includegraphics[width=0.31\linewidth]{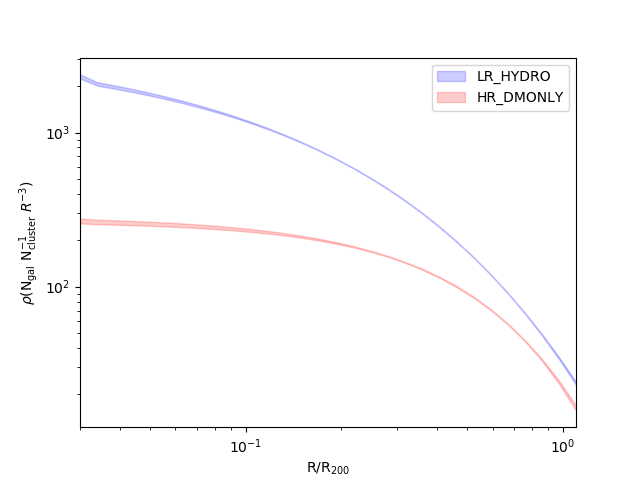}
    \includegraphics[width=0.68\linewidth]{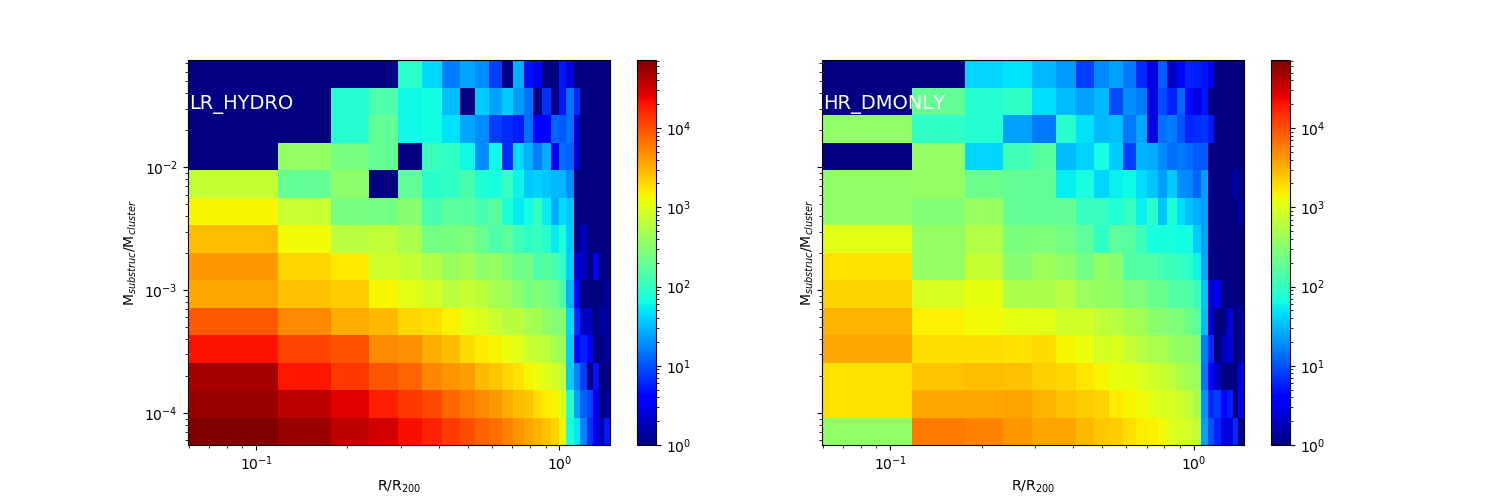}\hfil
    \includegraphics[width=0.31\linewidth]{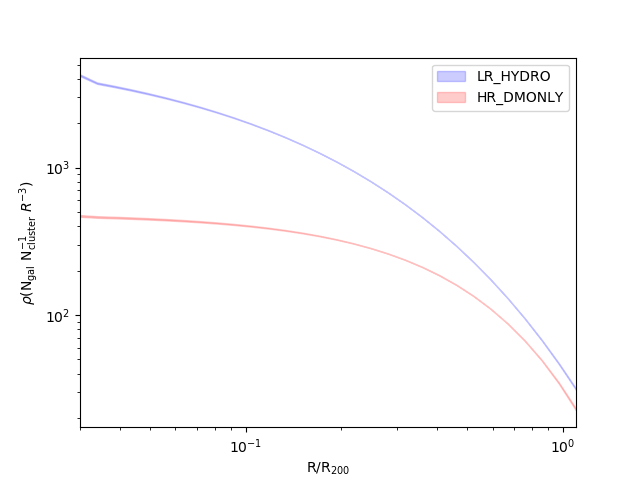}
    \includegraphics[width=0.68\linewidth]{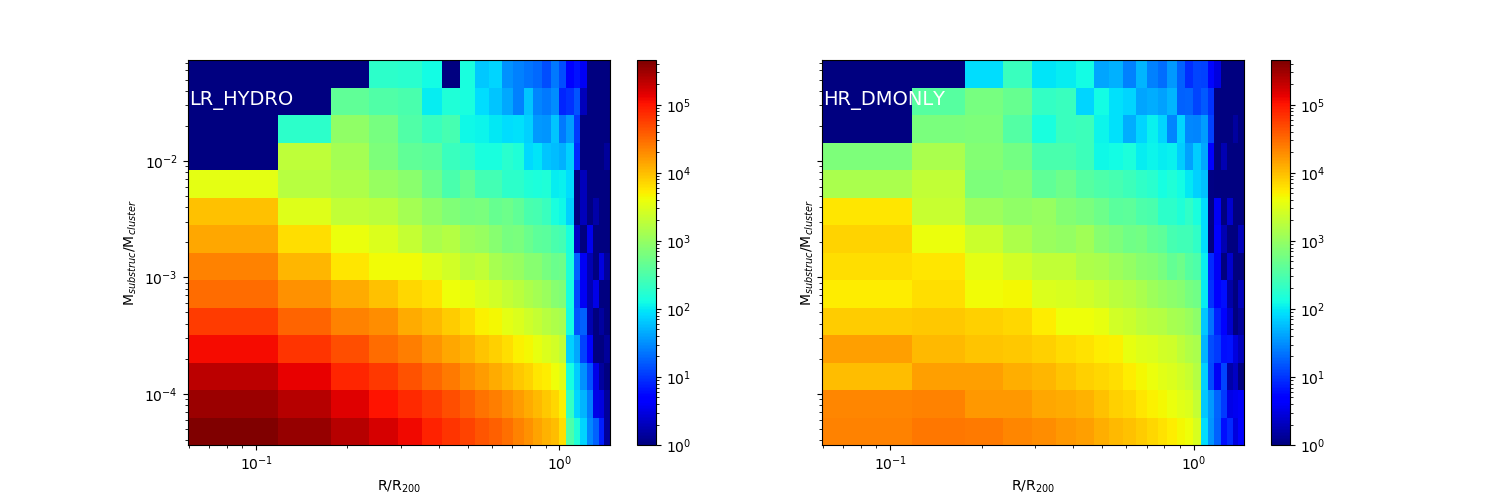}\hfil
\caption{3D radial number density of galaxies for the four bins in mass defined in Table~\ref{table_ch4:mass_bins} (from top to bottom) at redshift zero. The left column gives the best-fit Einasto model and uncertainties for the  \LRH\ (blue) and \HRD\ (red) simulations. In the middle and right column we display the radial galaxy density distribution per galaxy-cluster relative mass for the \LRH\ and \HRD\ simulations, respectively. The colorbars the number density of galaxies.\label{fig:radialdensityrelativemass} }

\end{figure*}

\subsection{Resolution Cuts on the Galaxy Mass Function}
\label{subsec_ch4:reso_cuts}

As discussed above the subhalo mass function is very similar for the four simulations flavors apart from resolution effects (see Fig~\ref{fig_ch4:all_subhalo}). The latter affect more severely to the \LRD\ simulations than to the \LRH\ ones. Comparing these two flavors in the following would require to apply a cut on the minimum mass of the galaxies considered to avoid resolution effects on the \LRD\ simulations. 
A better alternative is to compare \HRD\ and \LRH\ simulations by applying a cut derived from the latter.\\

For this we have computed for the \LRH\ simulations the minimum $ \left( M_{\rm{substruc}}/M_{\rm{halo}}\right) $ value for which resolution effects are not significant:
$$ \| N^{\rm{LR\_HYDRO} }/\max{ \left(N^{\rm{LR\_HYDRO} } \right)} - 1 \| > 0.1 $$
where $N^{\rm{LR\_HYDRO}}$ is the subhalo mass function for the \LRH\ simulations. The mass cut found are given in Table~\ref{tab:cuthydro} for each bin in mass and for each redshift slice. These mass cuts will be applied in the rest of this paper when comparing the properties of the \LRH\ and \HRD\ simulations.

\section{Galaxy Density Distribution}
\label{sec_ch4:galdens}

We study now the distribution of cluster galaxy members. For this, we concentrate on the \LRH\ and \HRD\ simulations to minimize resolution effects while preserving sufficient statistics (number of cluster regions). We apply the galaxy relative mass cuts presented in Table~\ref{tab:cuthydro}. This should ensure fair comparison both in terms of statistics and resolution. We stress that in most cases dark matter only simulations can be used to assess cluster detection algorithm performance \citep[see for example][]{adam2019euclid}.

\subsection{Galaxy number density radial profiles}
\label{subsec_ch4:methodology}


To assess the distribution of cluster galaxy members we compute galaxy density radial profiles for each of the identified clusters. We use equally spaced logarithmic radial bins in $R_{200}$ units so that we can easily compare and combine clusters of different masses and redshifts. We calculate the cumulative cluster galaxy member density from the center to the outskirt of the cluster. For a spherical shell at a distance $R/R_{200}$ from the cluster center, we count all the galaxies that are inside this sphere, and compute the associated spherical volume. We have also produced non cumulative radial profiles but found them less stable in terms of model fitting and will not be further discussed here. \\

We present in Figure~\ref{fig_ch4:HR_HYDRO_density_profile_fit} the cumulative galaxy number density profiles for the \LRH\ (left column) and \HRD\ (right column) simulations at redshift zero. From top to bottom we show results for the four bins in mass presented in Table~\ref{table_ch4:mass_bins}. The black dots correspond to stacked radial profiles. The stacked profiles are computed by counting all galaxies for all the clusters in the given radial bin and dividing by the volume and the number of clusters considered. Uncertainties are obtained assuming a Poisson distribution for the number of galaxies. We find that for the most inner radius the profiles are badly reconstructed due to a lack of statistics. This effect leads to a clear bias of the stacked galaxy density radial profiles as discussed in Appendix A (Figure~\ref{fig:radialdistributionallz}). In the following we will restrict the analysis to the external radial region as represented by the stacked profiles in Figure~\ref{fig_ch4:HR_HYDRO_density_profile_fit}.

\subsection{Modelling and Fitting}
\label{subsec_ch4:modelling}

It is difficult to extract physical information from a direct comparison of the galaxy density profiles. Better insights can be obtained from well known analytical models. At this respect, we have fitted the galaxy density profiles to an Einasto model \citep{einasto1965construction,navarro2004inner} defined as

\begin{equation}
    \rho(r) = n_0\exp{\left(\frac{-2}{\alpha}\left[  \left(\frac{r}{r_0}\right)^\alpha -1\right]\right)},
    \label{eq_ch4:einasto}
\end{equation}

\noindent where n$_0$, r$_0$ and $\alpha$ are free parameters. We summarize here the physical interpretation of the parameters of the model: 

\begin{itemize}

    \item[\textbullet] n$_0$ gives the normalization of the galaxy density.

    \item[\textbullet] $\alpha$ is related to the slope of the galaxy density distribution. When $\alpha$ decreases, the slope increases.
    
    \item[\textbullet] r$_0$ is a kind of characteristic radius. For a radius larger than r$_0$ the galaxy density profile drops rapidly. 
    
\end{itemize}

We have fitted the stacked cumulative galaxy density profiles to the Einasto model. Notice that the latter has been integrated to obtain a cumulative radial profile matching the procedure performed on the simulation data. For finding the best-fit parameters we have considered a MCMC approach based on the {\it emcee} python package \citep{2013PASP..125..306F}.  \\

We present in Figure~\ref{fig_ch4:HR_HYDRO_density_profile_fit} the results of the fit for the \LRH\ and \HRD\ simulations at redshift zero and for the four bin in mass. Overall the best-fit models are good fits to the data within the region of interest. The red lines correspond to the best-fit model for the stacked profiles in the case of the MCMC algorithms. We observe that the best-fit model represent a good fit to the data in the region of interest. The MCMC results also allowed us to handle the uncertainties on the parameters. \\

We show in Figures \ref{fig:einasto_alphar0} and \ref{fig:einasto_r0n0} the MCMC 2D probability distribution for the best-fit parameters of the stacked profiles in Figure~\ref{fig:radialdistributionallz}. For both figures we plot the 68 \% and 95.4 \% C.L. contours for \LRH\ (left column) and \HRD\ (right column) simulations. We give results for the four bins in mass in Table~\ref{table_ch4:mass_bins} (top to bottom) and for six slice in redshift from 0 to 1.4 (color coded). In some  cases and for the highest redshift the statistic is not sufficient to obtain reliable fits.

We observe in Fig.~\ref{fig:einasto_alphar0} a clear anti-correlation between the $\alpha$ and $r_{0}$ parameters. For both simulations the $r_0$ parameter decreases (smaller clusters) with increasing redshift as one would expect. Overall we observe larger values of the $\alpha$ and $r_{0}$ for the \HRD\ simulations with respect to the \LRH\ ones. This would indicate that the former are more concentrated and show steeper density profiles. Furthermore, for the \HRD\ simulations there seems to be a hint of redshift evolution of the $\alpha$ parameter with larger values at higher redshift. By contrast, for the \LRH\ simulations the $\alpha$ parameter shows very little evolution with redshift within the uncertainties. Nevertheless, it is difficult to give general conclusions on redshift evolution for any of the simulations as for a given bin in mass the distribution of cluster masses is very different across redshifts. Nevertheless, we conclude that as observed in Figures~\ref{fig_ch4:HR_HYDRO_density_profile_fit} and \ref{fig:radialdistributionallz} there are more galaxy in the inner cluster region for the \LRH\ simulations than for the \HRD\ ones. \\

The latter conclusions seem to be confirmed by the results shown in Figure~\ref{fig:einasto_r0n0} where we present the MCMC 2D probability distribution for the $r_{0}$  and $n_0$ parameters. We observe that the normalization parameter is larger for the \LRH\ simulations with respect to the \HRD\ ones. We also find that overall at lower redshift $n_0$ and $r_{0}$ are smaller and larger, respectively. This would indicate clusters are more extended at lower redshift as expected. As for the results in Fig.~\ref{fig:einasto_alphar0} the differences in the cluster mass distribution across redshift may explain some variations with respect to this general pattern. 

\subsection{Discussion}

From the previous results we can conclude that the density radial profile for \LRH\ simulations are more concentrated and {\it cuspier}, than those of the \HRD\ simulations. This can be clearly observed on the left column of Figure~\ref{fig:radialdensityrelativemass} where we present the best-fit Einasto galaxy density models and uncertainties (shaded region) for the \LRH\ (blue) and \HRD\ (red) simulations at redshift zero and for the four mass bins in Table~\ref{table_ch4:mass_bins} (top to bottom). The \LRH\ clusters present a clear excess of galaxies in the central region and a drop in the profile for smaller radius than the \HRD\ ones. In addition, we have seen from the comparison of the subhalo mass function in Section~\ref{subsed:massfunction} that the \LRH\ (\HRH) simulations have more low mass galaxies than the \LRD\ (\HRD) ones.
We interpret these results as the fact that baryons physics preserve low mass substructures that tend to locate in the center of the cluster. \\

We have checked the latter statement by studying the galaxy density as a function of the relative galaxy-cluster mass, $M_{\rm{substruc}}/M_{\rm{halo}}$. We show the results on the middle and right columns of Figure~\ref{fig:radialdensityrelativemass} at redshift zero for the \LRH\ and \HRD\ simulations, respectively. We represent the radial galaxy density (color coded) for different bins in $M_{\rm{substruc}}/M_{\rm{halo}}$. We confirm a clear excess of low mass galaxy in the inner cluster region for the \LRH\ simulations with respect to the \HRD\ ones. We also find that the low mass galaxies are present at all radius and are the main component in number density for both types of simulations. \\

\section{Conclusions}
\label{sec:conclu}

In this paper we have studied the properties of galaxies in the \TTH\ simulations in terms of luminosity function and radial distribution in the perspective of future large scale structure surveys at optical and infrared wavelengths. For this we have used full physics hydrodynamical simulations at at mass resolution of 1.5$\times$10$^9$ h$^{-1}$ M$_{\odot}$. We have completed these with equivalent resolution and eight times more resolved dark matter only simulations to disentangle possible effects from baryonic physics. 

In terms of the luminosity function we have found that the current mass resolution of the \TTH\ hydrodynamical simulations is not sufficient leading to artificial cut above magnitudes of about 21 when future surveys like Euclid is expected to go as deep as magnitudes of 24. These results have motivated the \TTH\ collaboration to produce five high resolution full-physics hydrodynamical regions (eight times more resolved in mass). We have proved that for those the reconstruction of the luminosity function goes up to magnitudes of 24. An effort to produce more high resolution hydro simulations is under way.  

We have computed subhalo (galaxies) mass function for the four flavor of \TTH\ simulations (hydro low and high resolution and dark matter high and low resolution) for four bins in mass and for six snapshots in redshift. We find that the lack of resolution leads to significantly less low mass galaxies both for the hydro and dark matter only simulations. We also find that the overall number of galaxies is the same between the low resolution hydro simulations and the high resolution dark matter only ones. We observe that overall the mass function are quite similar in the high mass regions where there are no resolution effects. We also find that baryonic physics tends to preserve significantly more low mass galaxies. We have approached the different subhalo mass functions by a power law and have consistent slopes across simulations flavors and no evolution with redshift. We conclude that the main difference between hydrodynamical and dark matter only simulations is the increased in the number of low mass galaxies. 

Finally, we have studied the radial distribution of cluster member galaxies. For study possible baryonic physic effects we compare the low resolution hydrodynamical simulations and the high resolution dark matter only resolution after imposing a cut in the relative mass of the particle for each bin in mass and for each redshift slice. Overall, we find that radial galaxy number density profiles of the hydro simulations show an excess of galaxies in the inner part of the cluster and are more concentrated than those of the dark matter only simulations. Furthermore, we observe that for the hydro simulations the low mass galaxies tend to concentrate on the inner part of the cluster. We conclude that baryonic physics preserves significantly more low mass substructures and in particular in the inner cluster region.

\bibliography{./references.bib}

\begin{thebibliography}{40}
\expandafter\ifx\csname natexlab\endcsname\relax\def\natexlab#1{#1}\fi

\bibitem[{{Abbott} {et~al.}(2020){Abbott}, {Aguena}, {Alarcon}, {Allam},
  {Allen}, {Annis}, {Avila}, {Bacon}, {Bechtol}, {Bermeo}, {Bernstein},
  {Bertin}, {Bhargava}, {Bocquet}, {Brooks}, {Brout}, {Buckley-Geer}, {Burke},
  {Carnero Rosell}, {Carrasco Kind}, {Carretero}, {Castander}, {Cawthon},
  {Chang}, {Chen}, {Choi}, {Costanzi}, {Crocce}, {da Costa}, {Davis}, {De
  Vicente}, {DeRose}, {Desai}, {Diehl}, {Dietrich}, {Dodelson}, {Doel},
  {Drlica-Wagner}, {Eckert}, {Eifler}, {Elvin-Poole}, {Estrada}, {Everett},
  {Evrard}, {Farahi}, {Ferrero}, {Flaugher}, {Fosalba}, {Frieman},
  {Garc{\'\i}a-Bellido}, {Gatti}, {Gaztanaga}, {Gerdes}, {Giannantonio},
  {Giles}, {Grandis}, {Gruen}, {Gruendl}, {Gschwend}, {Gutierrez}, {Hartley},
  {Hinton}, {Hollowood}, {Honscheid}, {Hoyle}, {Huterer}, {James}, {Jarvis},
  {Jeltema}, {Johnson}, {Johnson}, {Kent}, {Krause}, {Kron}, {Kuehn},
  {Kuropatkin}, {Lahav}, {Li}, {Lidman}, {Lima}, {Lin}, {MacCrann}, {Maia},
  {Mantz}, {Marshall}, {Martini}, {Mayers}, {Melchior}, {Mena-Fern{\'a}ndez},
  {Menanteau}, {Miquel}, {Mohr}, {Nichol}, {Nord}, {Ogando}, {Palmese},
  {Paz-Chinch{\'o}n}, {Plazas}, {Prat}, {Rau}, {Romer}, {Roodman}, {Rooney},
  {Rozo}, {Rykoff}, {Sako}, {Samuroff}, {S{\'a}nchez}, {Sanchez}, {Saro},
  {Scarpine}, {Schubnell}, {Scolnic}, {Serrano}, {Sevilla-Noarbe}, {Sheldon},
  {Smith}, {Smith}, {Suchyta}, {Swanson}, {Tarle}, {Thomas}, {To}, {Troxel},
  {Tucker}, {Varga}, {von der Linden}, {Walker}, {Wechsler}, {Weller},
  {Wilkinson}, {Wu}, {Yanny}, {Zhang}, {Zhang}, {Zuntz}, \& {DES
  Collaboration}}]{2020PhRvD.102b3509A}
{Abbott}, T.~M.~C., {Aguena}, M., {Alarcon}, A., {et~al.} 2020, \prd, 102,
  023509

\bibitem[{Adam {et~al.}(2019)Adam, Vannier, Maurogordato, Biviano, Adami,
  Ascaso, Bellagamba, Benoist, Cappi, D{\'\i}az-S{\'a}nchez,
  {et~al.}}]{adam2019euclid}
Adam, R., Vannier, M., Maurogordato, S., {et~al.} 2019, Astronomy \&
  Astrophysics, 627, A23

\bibitem[{{Adami} {et~al.}(2018){Adami}, {Giles}, {Koulouridis}, {Pacaud},
  {Caretta}, {Pierre}, {Eckert}, {Ramos-Ceja}, {Gastaldello}, {Fotopoulou},
  {Guglielmo}, {Lidman}, {Sadibekova}, {Iovino}, {Maughan}, {Chiappetti},
  {Alis}, {Altieri}, {Baldry}, {Bottini}, {Birkinshaw}, {Bremer}, {Brown},
  {Cucciati}, {Driver}, {Elmer}, {Ettori}, {Evrard}, {Faccioli}, {Granett},
  {Grootes}, {Guzzo}, {Hopkins}, {Horellou}, {Lef{\`e}vre}, {Liske}, {Malek},
  {Marulli}, {Maurogordato}, {Owers}, {Paltani}, {Poggianti}, {Polletta},
  {Plionis}, {Pollo}, {Pompei}, {Ponman}, {Rapetti}, {Ricci}, {Robotham},
  {Tuffs}, {Tasca}, {Valtchanov}, {Vergani}, {Wagner}, {Willis}, \& {XXL
  Consortium}}]{2018A&A...620A...5A}
{Adami}, C., {Giles}, P., {Koulouridis}, E., {et~al.} 2018, \aap, 620, A5

\bibitem[{{Allen} {et~al.}(2011){Allen}, {Evrard}, \&
  {Mantz}}]{2011ARA&A..49..409A}
{Allen}, S.~W., {Evrard}, A.~E., \& {Mantz}, A.~B. 2011, \araa, 49, 409

\bibitem[{Behroozi {et~al.}(2012)Behroozi, Wechsler, \&
  Wu}]{behroozi2012rockstar}
Behroozi, P.~S., Wechsler, R.~H., \& Wu, H.-Y. 2012, The Astrophysical Journal,
  762, 109

\bibitem[{Benitez(2000)}]{benitez2000bayesian}
Benitez, N. 2000, The Astrophysical Journal, 536, 571

\bibitem[{{Bleem} {et~al.}(2015){Bleem}, {Stalder}, {de Haan}, {Aird}, {Allen},
  {Applegate}, {Ashby}, {Bautz}, {Bayliss}, {Benson}, {Bocquet}, {Brodwin},
  {Carlstrom}, {Chang}, {Chiu}, {Cho}, {Clocchiatti}, {Crawford}, {Crites},
  {Desai}, {Dietrich}, {Dobbs}, {Foley}, {Forman}, {George}, {Gladders},
  {Gonzalez}, {Halverson}, {Hennig}, {Hoekstra}, {Holder}, {Holzapfel},
  {Hrubes}, {Jones}, {Keisler}, {Knox}, {Lee}, {Leitch}, {Liu}, {Lueker},
  {Luong-Van}, {Mantz}, {Marrone}, {McDonald}, {McMahon}, {Meyer}, {Mocanu},
  {Mohr}, {Murray}, {Padin}, {Pryke}, {Reichardt}, {Rest}, {Ruel}, {Ruhl},
  {Saliwanchik}, {Saro}, {Sayre}, {Schaffer}, {Schrabback}, {Shirokoff},
  {Song}, {Spieler}, {Stanford}, {Staniszewski}, {Stark}, {Story}, {Stubbs},
  {Vanderlinde}, {Vieira}, {Vikhlinin}, {Williamson}, {Zahn}, \&
  {Zenteno}}]{2015ApJS..216...27B}
{Bleem}, L.~E., {Stalder}, B., {de Haan}, T., {et~al.} 2015, \apjs, 216, 27

\bibitem[{{Bocquet} {et~al.}(2019){Bocquet}, {Dietrich}, {Schrabback}, {Bleem},
  {Klein}, {Allen}, {Applegate}, {Ashby}, {Bautz}, {Bayliss}, {Benson},
  {Brodwin}, {Bulbul}, {Canning}, {Capasso}, {Carlstrom}, {Chang}, {Chiu},
  {Cho}, {Clocchiatti}, {Crawford}, {Crites}, {de Haan}, {Desai}, {Dobbs},
  {Foley}, {Forman}, {Garmire}, {George}, {Gladders}, {Gonzalez}, {Grandis},
  {Gupta}, {Halverson}, {Hlavacek-Larrondo}, {Hoekstra}, {Holder}, {Holzapfel},
  {Hou}, {Hrubes}, {Huang}, {Jones}, {Khullar}, {Knox}, {Kraft}, {Lee}, {von
  der Linden}, {Luong-Van}, {Mantz}, {Marrone}, {McDonald}, {McMahon}, {Meyer},
  {Mocanu}, {Mohr}, {Morris}, {Padin}, {Patil}, {Pryke}, {Rapetti},
  {Reichardt}, {Rest}, {Ruhl}, {Saliwanchik}, {Saro}, {Sayre}, {Schaffer},
  {Shirokoff}, {Stalder}, {Stanford}, {Staniszewski}, {Stark}, {Story},
  {Strazzullo}, {Stubbs}, {Vanderlinde}, {Vieira}, {Vikhlinin}, {Williamson},
  \& {Zenteno}}]{2019ApJ...878...55B}
{Bocquet}, S., {Dietrich}, J.~P., {Schrabback}, T., {et~al.} 2019, \apj, 878,
  55

\bibitem[{{B{\"o}hringer} {et~al.}(2017){B{\"o}hringer}, {Chon}, \&
  {Fukugita}}]{2017A&A...608A..65B}
{B{\"o}hringer}, H., {Chon}, G., \& {Fukugita}, M. 2017, \aap, 608, A65

\bibitem[{Chabrier(2003)}]{chabrier2003galactic}
Chabrier, G. 2003, Publications of the Astronomical Society of the Pacific,
  115, 763

\bibitem[{{Cui} {et~al.}(2022){Cui}, {Dave}, {Knebe}, {Rasia}, {Gray},
  {Pearce}, {Power}, {Yepes}, {Anbajagane}, {Ceverino}, {Contreras-Santos}, {de
  Andres}, {De Petris}, {Ettori}, {Haggar}, {Li}, {Wang}, {Yang}, {Borgani},
  {Dolag}, {Zu}, {Kuchner}, {Ca{\~n}as}, {Ferragamo}, \&
  {Gianfagna}}]{2022MNRAS.514..977C}
{Cui}, W., {Dave}, R., {Knebe}, A., {et~al.} 2022, \mnras, 514, 977

\bibitem[{Cui {et~al.}(2018)Cui, Knebe, Yepes, Pearce, Power, Dave, Arth,
  Borgani, Dolag, Elahi, {et~al.}}]{cui2018three}
Cui, W., Knebe, A., Yepes, G., {et~al.} 2018, Monthly Notices of the Royal
  Astronomical Society, 480, 2898

\bibitem[{{Dav{\'e}} {et~al.}(2019){Dav{\'e}}, {Angl{\'e}s-Alc{\'a}zar},
  {Narayanan}, {Li}, {Rafieferantsoa}, \& {Appleby}}]{2019MNRAS.486.2827D}
{Dav{\'e}}, R., {Angl{\'e}s-Alc{\'a}zar}, D., {Narayanan}, D., {et~al.} 2019,
  \mnras, 486, 2827

\bibitem[{{Dav{\'e}} {et~al.}(2016){Dav{\'e}}, {Thompson}, \&
  {Hopkins}}]{2016MNRAS.462.3265D}
{Dav{\'e}}, R., {Thompson}, R., \& {Hopkins}, P.~F. 2016, \mnras, 462, 3265

\bibitem[{{de Andres} {et~al.}(2023){de Andres}, {Yepes}, {Sembolini},
  {Mart{\'\i}nez-Mu{\~n}oz}, {Cui}, {Robledo}, {Chuang}, \&
  {Rasia}}]{2023MNRAS.518..111D}
{de Andres}, D., {Yepes}, G., {Sembolini}, F., {et~al.} 2023, \mnras, 518, 111

\bibitem[{{de Haan} {et~al.}(2016){de Haan}, {Benson}, {Bleem}, {Allen},
  {Applegate}, {Ashby}, {Bautz}, {Bayliss}, {Bocquet}, {Brodwin}, {Carlstrom},
  {Chang}, {Chiu}, {Cho}, {Clocchiatti}, {Crawford}, {Crites}, {Desai},
  {Dietrich}, {Dobbs}, {Doucouliagos}, {Foley}, {Forman}, {Garmire}, {George},
  {Gladders}, {Gonzalez}, {Gupta}, {Halverson}, {Hlavacek-Larrondo},
  {Hoekstra}, {Holder}, {Holzapfel}, {Hou}, {Hrubes}, {Huang}, {Jones},
  {Keisler}, {Knox}, {Lee}, {Leitch}, {von der Linden}, {Luong-Van}, {Mantz},
  {Marrone}, {McDonald}, {McMahon}, {Meyer}, {Mocanu}, {Mohr}, {Murray},
  {Padin}, {Pryke}, {Rapetti}, {Reichardt}, {Rest}, {Ruel}, {Ruhl},
  {Saliwanchik}, {Saro}, {Sayre}, {Schaffer}, {Schrabback}, {Shirokoff},
  {Song}, {Spieler}, {Stalder}, {Stanford}, {Staniszewski}, {Stark}, {Story},
  {Stubbs}, {Vanderlinde}, {Vieira}, {Vikhlinin}, {Williamson}, \&
  {Zenteno}}]{2016ApJ...832...95D}
{de Haan}, T., {Benson}, B.~A., {Bleem}, L.~E., {et~al.} 2016, \apj, 832, 95

\bibitem[{{Devriendt} {et~al.}(1999){Devriendt}, {Guiderdoni}, \&
  {Sadat}}]{1999A&A...350..381D}
{Devriendt}, J.~E.~G., {Guiderdoni}, B., \& {Sadat}, R. 1999, \aap, 350, 381

\bibitem[{Dolag {et~al.}(2009)Dolag, Borgani, Murante, \&
  Springel}]{dolag2009substructures}
Dolag, K., Borgani, S., Murante, G., \& Springel, V. 2009, Monthly Notices of
  the Royal Astronomical Society, 399, 497

\bibitem[{{Drlica-Wagner} {et~al.}(2018){Drlica-Wagner}, {Sevilla-Noarbe},
  {Rykoff}, {Gruendl}, {Yanny}, {Tucker}, {Hoyle}, {Carnero Rosell},
  {Bernstein}, {Bechtol}, {Becker}, {Benoit-L{\'e}vy}, {Bertin}, {Carrasco
  Kind}, {Davis}, {de Vicente}, {Diehl}, {Gruen}, {Hartley}, {Leistedt}, {Li},
  {Marshall}, {Neilsen}, {Rau}, {Sheldon}, {Smith}, {Troxel}, {Wyatt}, {Zhang},
  {Abbott}, {Abdalla}, {Allam}, {Banerji}, {Brooks}, {Buckley-Geer}, {Burke},
  {Capozzi}, {Carretero}, {Cunha}, {D'Andrea}, {da Costa}, {DePoy}, {Desai},
  {Dietrich}, {Doel}, {Evrard}, {Fausti Neto}, {Flaugher}, {Fosalba},
  {Frieman}, {Garc{\'\i}a-Bellido}, {Gerdes}, {Giannantonio}, {Gschwend},
  {Gutierrez}, {Honscheid}, {James}, {Jeltema}, {Kuehn}, {Kuhlmann},
  {Kuropatkin}, {Lahav}, {Lima}, {Lin}, {Maia}, {Martini}, {McMahon},
  {Melchior}, {Menanteau}, {Miquel}, {Nichol}, {Ogando}, {Plazas}, {Romer},
  {Roodman}, {Sanchez}, {Scarpine}, {Schindler}, {Schubnell}, {Smith}, {Smith},
  {Soares-Santos}, {Sobreira}, {Suchyta}, {Tarle}, {Vikram}, {Walker},
  {Wechsler}, {Zuntz}, \& {DES Collaboration}}]{2018ApJS..235...33D}
{Drlica-Wagner}, A., {Sevilla-Noarbe}, I., {Rykoff}, E.~S., {et~al.} 2018,
  \apjs, 235, 33

\bibitem[{Einasto(1965)}]{einasto1965construction}
Einasto, J. 1965, Trudy Astrofizicheskogo Instituta Alma-Ata, 5, 87

\bibitem[{{Foreman-Mackey} {et~al.}(2013){Foreman-Mackey}, {Hogg}, {Lang}, \&
  {Goodman}}]{2013PASP..125..306F}
{Foreman-Mackey}, D., {Hogg}, D.~W., {Lang}, D., \& {Goodman}, J. 2013, \pasp,
  125, 306

\bibitem[{{Haardt} \& {Madau}(2012)}]{2012ApJ...746..125H}
{Haardt}, F. \& {Madau}, P. 2012, \apj, 746, 125

\bibitem[{{Hilton} {et~al.}(2021){Hilton}, {Sif{\'o}n}, {Naess},
  {Madhavacheril}, {Oguri}, {Rozo}, {Rykoff}, {Abbott}, {Adhikari}, {Aguena},
  {Aiola}, {Allam}, {Amodeo}, {Amon}, {Annis}, {Ansarinejad}, {Aros-Bunster},
  {Austermann}, {Avila}, {Bacon}, {Battaglia}, {Beall}, {Becker}, {Bernstein},
  {Bertin}, {Bhandarkar}, {Bhargava}, {Bond}, {Brooks}, {Burke}, {Calabrese},
  {Carrasco Kind}, {Carretero}, {Choi}, {Choi}, {Conselice}, {da Costa},
  {Costanzi}, {Crichton}, {Crowley}, {D{\"u}nner}, {Denison}, {Devlin},
  {Dicker}, {Diehl}, {Dietrich}, {Doel}, {Duff}, {Duivenvoorden}, {Dunkley},
  {Everett}, {Ferraro}, {Ferrero}, {Fert{\'e}}, {Flaugher}, {Frieman},
  {Gallardo}, {Garc{\'\i}a-Bellido}, {Gaztanaga}, {Gerdes}, {Giles}, {Golec},
  {Gralla}, {Grandis}, {Gruen}, {Gruendl}, {Gschwend}, {Gutierrez}, {Han},
  {Hartley}, {Hasselfield}, {Hill}, {Hilton}, {Hincks}, {Hinton}, {Ho},
  {Honscheid}, {Hoyle}, {Hubmayr}, {Huffenberger}, {Hughes}, {Jaelani}, {Jain},
  {James}, {Jeltema}, {Kent}, {Knowles}, {Koopman}, {Kuehn}, {Lahav}, {Lima},
  {Lin}, {Lokken}, {Loubser}, {MacCrann}, {Maia}, {Marriage}, {Martin},
  {McMahon}, {Melchior}, {Menanteau}, {Miquel}, {Miyatake}, {Moodley},
  {Morgan}, {Mroczkowski}, {Nati}, {Newburgh}, {Niemack}, {Nishizawa},
  {Ogando}, {Orlowski-Scherer}, {Page}, {Palmese}, {Partridge},
  {Paz-Chinch{\'o}n}, {Phakathi}, {Plazas}, {Robertson}, {Romer}, {Carnero
  Rosell}, {Salatino}, {Sanchez}, {Schaan}, {Schillaci}, {Sehgal}, {Serrano},
  {Shin}, {Simon}, {Smith}, {Soares-Santos}, {Spergel}, {Staggs}, {Storer},
  {Suchyta}, {Swanson}, {Tarle}, {Thomas}, {To}, {Trac}, {Ullom}, {Vale}, {Van
  Lanen}, {Vavagiakis}, {De Vicente}, {Wilkinson}, {Wollack}, {Xu}, \&
  {Zhang}}]{2021ApJS..253....3H}
{Hilton}, M., {Sif{\'o}n}, C., {Naess}, S., {et~al.} 2021, \apjs, 253, 3

\bibitem[{{Hopkins}(2015)}]{2015MNRAS.450...53H}
{Hopkins}, P.~F. 2015, \mnras, 450, 53

\bibitem[{{Klypin} {et~al.}(2016){Klypin}, {Yepes}, {Gottl{\"o}ber}, {Prada},
  \& {He{\ss}}}]{2016MNRAS.457.4340K}
{Klypin}, A., {Yepes}, G., {Gottl{\"o}ber}, S., {Prada}, F., \& {He{\ss}}, S.
  2016, \mnras, 457, 4340

\bibitem[{Knollmann \& Knebe(2009)}]{knollmann2009ahf}
Knollmann, S.~R. \& Knebe, A. 2009, The Astrophysical Journal Supplement
  Series, 182, 608

\bibitem[{{Laureijs} {et~al.}(2011){Laureijs}, {Amiaux}, {Arduini},
  {Augu{\`e}res}, {Brinchmann}, {Cole}, {Cropper}, {Dabin}, {Duvet}, {Ealet},
  {Garilli}, {Gondoin}, {Guzzo}, {Hoar}, {Hoekstra}, {Holmes}, {Kitching},
  {Maciaszek}, {Mellier}, {Pasian}, {Percival}, {Rhodes}, {Saavedra Criado},
  {Sauvage}, {Scaramella}, {Valenziano}, {Warren}, {Bender}, {Castander},
  {Cimatti}, {Le F{\`e}vre}, {Kurki-Suonio}, {Levi}, {Lilje}, {Meylan},
  {Nichol}, {Pedersen}, {Popa}, {Rebolo Lopez}, {Rix}, {Rottgering},
  {Zeilinger}, {Grupp}, {Hudelot}, {Massey}, {Meneghetti}, {Miller}, {Paltani},
  {Paulin-Henriksson}, {Pires}, {Saxton}, {Schrabback}, {Seidel}, {Walsh},
  {Aghanim}, {Amendola}, {Bartlett}, {Baccigalupi}, {Beaulieu}, {Benabed},
  {Cuby}, {Elbaz}, {Fosalba}, {Gavazzi}, {Helmi}, {Hook}, {Irwin}, {Kneib},
  {Kunz}, {Mannucci}, {Moscardini}, {Tao}, {Teyssier}, {Weller}, {Zamorani},
  {Zapatero Osorio}, {Boulade}, {Foumond}, {Di Giorgio}, {Guttridge}, {James},
  {Kemp}, {Martignac}, {Spencer}, {Walton}, {Bl{\"u}mchen}, {Bonoli},
  {Bortoletto}, {Cerna}, {Corcione}, {Fabron}, {Jahnke}, {Ligori}, {Madrid},
  {Martin}, {Morgante}, {Pamplona}, {Prieto}, {Riva}, {Toledo}, {Trifoglio},
  {Zerbi}, {Abdalla}, {Douspis}, {Grenet}, {Borgani}, {Bouwens}, {Courbin},
  {Delouis}, {Dubath}, {Fontana}, {Frailis}, {Grazian}, {Koppenh{\"o}fer},
  {Mansutti}, {Melchior}, {Mignoli}, {Mohr}, {Neissner}, {Noddle}, {Poncet},
  {Scodeggio}, {Serrano}, {Shane}, {Starck}, {Surace}, {Taylor},
  {Verdoes-Kleijn}, {Vuerli}, {Williams}, {Zacchei}, {Altieri}, {Escudero
  Sanz}, {Kohley}, {Oosterbroek}, {Astier}, {Bacon}, {Bardelli}, {Baugh},
  {Bellagamba}, {Benoist}, {Bianchi}, {Biviano}, {Branchini}, {Carbone},
  {Cardone}, {Clements}, {Colombi}, {Conselice}, {Cresci}, {Deacon}, {Dunlop},
  {Fedeli}, {Fontanot}, {Franzetti}, {Giocoli}, {Garcia-Bellido}, {Gow},
  {Heavens}, {Hewett}, {Heymans}, {Holland}, {Huang}, {Ilbert}, {Joachimi},
  {Jennins}, {Kerins}, {Kiessling}, {Kirk}, {Kotak}, {Krause}, {Lahav}, {van
  Leeuwen}, {Lesgourgues}, {Lombardi}, {Magliocchetti}, {Maguire}, {Majerotto},
  {Maoli}, {Marulli}, {Maurogordato}, {McCracken}, {McLure}, {Melchiorri},
  {Merson}, {Moresco}, {Nonino}, {Norberg}, {Peacock}, {Pello}, {Penny},
  {Pettorino}, {Di Porto}, {Pozzetti}, {Quercellini}, {Radovich}, {Rassat},
  {Roche}, {Ronayette}, {Rossetti}, {Sartoris}, {Schneider}, {Semboloni},
  {Serjeant}, {Simpson}, {Skordis}, {Smadja}, {Smartt}, {Spano}, {Spiro},
  {Sullivan}, {Tilquin}, {Trotta}, {Verde}, {Wang}, {Williger}, {Zhao},
  {Zoubian}, \& {Zucca}}]{2011arXiv1110.3193L}
{Laureijs}, R., {Amiaux}, J., {Arduini}, S., {et~al.} 2011, arXiv e-prints,
  arXiv:1110.3193

\bibitem[{{Liu} {et~al.}(2022){Liu}, {Bulbul}, {Ghirardini}, {Liu}, {Klein},
  {Clerc}, {{\"O}zsoy}, {Ramos-Ceja}, {Pacaud}, {Comparat}, {Okabe}, {Bahar},
  {Biffi}, {Brunner}, {Br{\"u}ggen}, {Buchner}, {Ider Chitham}, {Chiu},
  {Dolag}, {Gatuzz}, {Gonzalez}, {Hoang}, {Lamer}, {Merloni}, {Nandra},
  {Oguri}, {Ota}, {Predehl}, {Reiprich}, {Salvato}, {Schrabback}, {Sanders},
  {Seppi}, \& {Thibaud}}]{2022A&A...661A...2L}
{Liu}, A., {Bulbul}, E., {Ghirardini}, V., {et~al.} 2022, \aap, 661, A2

\bibitem[{{Maturi} {et~al.}(2019){Maturi}, {Bellagamba}, {Radovich},
  {Roncarelli}, {Sereno}, {Moscardini}, {Bardelli}, \&
  {Puddu}}]{2019MNRAS.485..498M}
{Maturi}, M., {Bellagamba}, F., {Radovich}, M., {et~al.} 2019, \mnras, 485, 498

\bibitem[{Navarro {et~al.}(2004)Navarro, Hayashi, Power, Jenkins, Frenk, White,
  Springel, Stadel, \& Quinn}]{navarro2004inner}
Navarro, J.~F., Hayashi, E., Power, C., {et~al.} 2004, Monthly Notices of the
  Royal Astronomical Society, 349, 1039

\bibitem[{{Pacaud} {et~al.}(2018){Pacaud}, {Pierre}, {Melin}, {Adami},
  {Evrard}, {Galli}, {Gastaldello}, {Maughan}, {Sereno}, {Alis}, {Altieri},
  {Birkinshaw}, {Chiappetti}, {Faccioli}, {Giles}, {Horellou}, {Iovino},
  {Koulouridis}, {Le F{\`e}vre}, {Lidman}, {Lieu}, {Maurogordato},
  {Moscardini}, {Plionis}, {Poggianti}, {Pompei}, {Sadibekova}, {Valtchanov},
  \& {Willis}}]{2018A&A...620A..10P}
{Pacaud}, F., {Pierre}, M., {Melin}, J.~B., {et~al.} 2018, \aap, 620, A10

\bibitem[{{Planck Collaboration} {et~al.}(2014){Planck Collaboration}, {Ade},
  {Aghanim}, {Armitage-Caplan}, {Arnaud}, {Ashdown}, {Atrio-Barandela},
  {Aumont}, {Baccigalupi}, {Banday}, {Barreiro}, {Barrena}, {Bartlett},
  {Battaner}, {Battye}, {Benabed}, {Beno{\^\i}t}, {Benoit-L{\'e}vy}, {Bernard},
  {Bersanelli}, {Bielewicz}, {Bikmaev}, {Blanchard}, {Bobin}, {Bock},
  {B{\"o}hringer}, {Bonaldi}, {Bond}, {Borrill}, {Bouchet}, {Bourdin},
  {Bridges}, {Brown}, {Bucher}, {Burenin}, {Burigana}, {Butler}, {Cardoso},
  {Carvalho}, {Catalano}, {Challinor}, {Chamballu}, {Chary}, {Chiang},
  {Chiang}, {Chon}, {Christensen}, {Church}, {Clements}, {Colombi}, {Colombo},
  {Couchot}, {Coulais}, {Crill}, {Curto}, {Cuttaia}, {Da Silva}, {Dahle},
  {Danese}, {Davies}, {Davis}, {de Bernardis}, {de Rosa}, {de Zotti},
  {Delabrouille}, {Delouis}, {D{\'e}mocl{\`e}s}, {D{\'e}sert}, {Dickinson},
  {Diego}, {Dolag}, {Dole}, {Donzelli}, {Dor{\'e}}, {Douspis}, {Dupac},
  {Efstathiou}, {En{\ss}lin}, {Eriksen}, {Finelli}, {Flores-Cacho}, {Forni},
  {Frailis}, {Franceschi}, {Fromenteau}, {Galeotta}, {Ganga},
  {G{\'e}nova-Santos}, {Giard}, {Giardino}, {Giraud-H{\'e}raud},
  {Gonz{\'a}lez-Nuevo}, {G{\'o}rski}, {Gratton}, {Gregorio}, {Gruppuso},
  {Hansen}, {Hanson}, {Harrison}, {Henrot-Versill{\'e}},
  {Hern{\'a}ndez-Monteagudo}, {Herranz}, {Hildebrandt}, {Hivon}, {Hobson},
  {Holmes}, {Hornstrup}, {Hovest}, {Huffenberger}, {Hurier}, {Jaffe}, {Jaffe},
  {Jones}, {Juvela}, {Keih{\"a}nen}, {Keskitalo}, {Khamitov}, {Kisner},
  {Kneissl}, {Knoche}, {Knox}, {Kunz}, {Kurki-Suonio}, {Lagache},
  {L{\"a}hteenm{\"a}ki}, {Lamarre}, {Lasenby}, {Laureijs}, {Lawrence}, {Leahy},
  {Leonardi}, {Le{\'o}n-Tavares}, {Lesgourgues}, {Liddle}, {Liguori}, {Lilje},
  {Linden-V{\o}rnle}, {L{\'o}pez-Caniego}, {Lubin}, {Mac{\'\i}as-P{\'e}rez},
  {Maffei}, {Maino}, {Mandolesi}, {Marcos-Caballero}, {Maris}, {Marshall},
  {Martin}, {Mart{\'\i}nez-Gonz{\'a}lez}, {Masi}, {Matarrese}, {Matthai},
  {Mazzotta}, {Meinhold}, {Melchiorri}, {Melin}, {Mendes}, {Mennella},
  {Migliaccio}, {Mitra}, {Miville-Desch{\^e}nes}, {Moneti}, {Montier},
  {Morgante}, {Mortlock}, {Moss}, {Munshi}, {Naselsky}, {Nati}, {Natoli},
  {Netterfield}, {N{\o}rgaard-Nielsen}, {Noviello}, {Novikov}, {Novikov},
  {Osborne}, {Oxborrow}, {Paci}, {Pagano}, {Pajot}, {Paoletti}, {Partridge},
  {Pasian}, {Patanchon}, {Perdereau}, {Perotto}, {Perrotta}, {Piacentini},
  {Piat}, {Pierpaoli}, {Pietrobon}, {Plaszczynski}, {Pointecouteau}, {Polenta},
  {Ponthieu}, {Popa}, {Poutanen}, {Pratt}, {Pr{\'e}zeau}, {Prunet}, {Puget},
  {Rachen}, {Rebolo}, {Reinecke}, {Remazeilles}, {Renault}, {Ricciardi},
  {Riller}, {Ristorcelli}, {Rocha}, {Roman}, {Rosset}, {Roudier},
  {Rowan-Robinson}, {Rubi{\~n}o-Mart{\'\i}n}, {Rusholme}, {Sandri}, {Santos},
  {Savini}, {Scott}, {Seiffert}, {Shellard}, {Spencer}, {Starck}, {Stolyarov},
  {Stompor}, {Sudiwala}, {Sunyaev}, {Sureau}, {Sutton}, {Suur-Uski}, {Sygnet},
  {Tauber}, {Tavagnacco}, {Terenzi}, {Toffolatti}, {Tomasi}, {Tristram},
  {Tucci}, {Tuovinen}, {T{\"u}rler}, {Umana}, {Valenziano}, {Valiviita}, {Van
  Tent}, {Vielva}, {Villa}, {Vittorio}, {Wade}, {Wandelt}, {Weller}, {White},
  {White}, {Yvon}, {Zacchei}, \& {Zonca}}]{2014A&A...571A..20P}
{Planck Collaboration}, {Ade}, P.~A.~R., {Aghanim}, N., {et~al.} 2014, \aap,
  571, A20

\bibitem[{{Planck Collaboration} {et~al.}(2016){Planck Collaboration}, {Ade},
  {Aghanim}, {Arnaud}, {Ashdown}, {Aumont}, {Baccigalupi}, {Banday},
  {Barreiro}, {Bartlett}, {Bartolo}, {Battaner}, {Battye}, {Benabed},
  {Beno{\^\i}t}, {Benoit-L{\'e}vy}, {Bernard}, {Bersanelli}, {Bielewicz},
  {Bock}, {Bonaldi}, {Bonavera}, {Bond}, {Borrill}, {Bouchet}, {Bucher},
  {Burigana}, {Butler}, {Calabrese}, {Cardoso}, {Catalano}, {Challinor},
  {Chamballu}, {Chary}, {Chiang}, {Christensen}, {Church}, {Clements},
  {Colombi}, {Colombo}, {Combet}, {Comis}, {Couchot}, {Coulais}, {Crill},
  {Curto}, {Cuttaia}, {Danese}, {Davies}, {Davis}, {de Bernardis}, {de Rosa},
  {de Zotti}, {Delabrouille}, {D{\'e}sert}, {Diego}, {Dolag}, {Dole},
  {Donzelli}, {Dor{\'e}}, {Douspis}, {Ducout}, {Dupac}, {Efstathiou}, {Elsner},
  {En{\ss}lin}, {Eriksen}, {Falgarone}, {Fergusson}, {Finelli}, {Forni},
  {Frailis}, {Fraisse}, {Franceschi}, {Frejsel}, {Galeotta}, {Galli}, {Ganga},
  {Giard}, {Giraud-H{\'e}raud}, {Gjerl{\o}w}, {Gonz{\'a}lez-Nuevo},
  {G{\'o}rski}, {Gratton}, {Gregorio}, {Gruppuso}, {Gudmundsson}, {Hansen},
  {Hanson}, {Harrison}, {Henrot-Versill{\'e}}, {Hern{\'a}ndez-Monteagudo},
  {Herranz}, {Hildebrandt}, {Hivon}, {Hobson}, {Holmes}, {Hornstrup}, {Hovest},
  {Huffenberger}, {Hurier}, {Jaffe}, {Jaffe}, {Jones}, {Juvela},
  {Keih{\"a}nen}, {Keskitalo}, {Kisner}, {Kneissl}, {Knoche}, {Kunz},
  {Kurki-Suonio}, {Lagache}, {L{\"a}hteenm{\"a}ki}, {Lamarre}, {Lasenby},
  {Lattanzi}, {Lawrence}, {Leonardi}, {Lesgourgues}, {Levrier}, {Liguori},
  {Lilje}, {Linden-V{\o}rnle}, {L{\'o}pez-Caniego}, {Lubin},
  {Mac{\'\i}as-P{\'e}rez}, {Maggio}, {Maino}, {Mandolesi}, {Mangilli}, {Maris},
  {Martin}, {Mart{\'\i}nez-Gonz{\'a}lez}, {Masi}, {Matarrese}, {McGehee},
  {Meinhold}, {Melchiorri}, {Melin}, {Mendes}, {Mennella}, {Migliaccio},
  {Mitra}, {Miville-Desch{\^e}nes}, {Moneti}, {Montier}, {Morgante},
  {Mortlock}, {Moss}, {Munshi}, {Murphy}, {Naselsky}, {Nati}, {Natoli},
  {Netterfield}, {N{\o}rgaard-Nielsen}, {Noviello}, {Novikov}, {Novikov},
  {Oxborrow}, {Paci}, {Pagano}, {Pajot}, {Paoletti}, {Partridge}, {Pasian},
  {Patanchon}, {Pearson}, {Perdereau}, {Perotto}, {Perrotta}, {Pettorino},
  {Piacentini}, {Piat}, {Pierpaoli}, {Pietrobon}, {Plaszczynski},
  {Pointecouteau}, {Polenta}, {Popa}, {Pratt}, {Pr{\'e}zeau}, {Prunet},
  {Puget}, {Rachen}, {Rebolo}, {Reinecke}, {Remazeilles}, {Renault}, {Renzi},
  {Ristorcelli}, {Rocha}, {Roman}, {Rosset}, {Rossetti}, {Roudier},
  {Rubi{\~n}o-Mart{\'\i}n}, {Rusholme}, {Sandri}, {Santos}, {Savelainen},
  {Savini}, {Scott}, {Seiffert}, {Shellard}, {Spencer}, {Stolyarov}, {Stompor},
  {Sudiwala}, {Sunyaev}, {Sutton}, {Suur-Uski}, {Sygnet}, {Tauber}, {Terenzi},
  {Toffolatti}, {Tomasi}, {Tristram}, {Tucci}, {Tuovinen}, {T{\"u}rler},
  {Umana}, {Valenziano}, {Valiviita}, {Van Tent}, {Vielva}, {Villa}, {Wade},
  {Wandelt}, {Wehus}, {Weller}, {White}, {Yvon}, {Zacchei}, \&
  {Zonca}}]{2016A&A...594A..24P}
{Planck Collaboration}, {Ade}, P.~A.~R., {Aghanim}, N., {et~al.} 2016, \aap,
  594, A24

\bibitem[{{Pratt} {et~al.}(2019){Pratt}, {Arnaud}, {Biviano}, {Eckert},
  {Ettori}, {Nagai}, {Okabe}, \& {Reiprich}}]{pratt2019}
{Pratt}, G.~W., {Arnaud}, M., {Biviano}, A., {et~al.} 2019, \ssr, 215, 25

\bibitem[{Press \& Schechter(1974)}]{press1974formation}
Press, W.~H. \& Schechter, P. 1974, The Astrophysical Journal, 187, 425

\bibitem[{{Rahmati} {et~al.}(2013){Rahmati}, {Pawlik}, {Rai{\v{c}}evi{\'c}}, \&
  {Schaye}}]{2013MNRAS.430.2427R}
{Rahmati}, A., {Pawlik}, A.~H., {Rai{\v{c}}evi{\'c}}, M., \& {Schaye}, J. 2013,
  \mnras, 430, 2427

\bibitem[{{Rykoff} {et~al.}(2014){Rykoff}, {Rozo}, {Busha}, {Cunha},
  {Finoguenov}, {Evrard}, {Hao}, {Koester}, {Leauthaud}, {Nord}, {Pierre},
  {Reddick}, {Sadibekova}, {Sheldon}, \& {Wechsler}}]{2014ApJ...785..104R}
{Rykoff}, E.~S., {Rozo}, E., {Busha}, M.~T., {et~al.} 2014, \apj, 785, 104

\bibitem[{{Sartoris} {et~al.}(2016){Sartoris}, {Biviano}, {Fedeli}, {Bartlett},
  {Borgani}, {Costanzi}, {Giocoli}, {Moscardini}, {Weller}, {Ascaso},
  {Bardelli}, {Maurogordato}, \& {Viana}}]{2016MNRAS.459.1764S}
{Sartoris}, B., {Biviano}, A., {Fedeli}, C., {et~al.} 2016, \mnras, 459, 1764

\bibitem[{{Smith} {et~al.}(2017){Smith}, {Bryan}, {Glover}, {Goldbaum}, {Turk},
  {Regan}, {Wise}, {Schive}, {Abel}, {Emerick}, {O'Shea}, {Anninos}, {Hummels},
  \& {Khochfar}}]{2017MNRAS.466.2217S}
{Smith}, B.~D., {Bryan}, G.~L., {Glover}, S. C.~O., {et~al.} 2017, \mnras, 466,
  2217

\bibitem[{Tinker {et~al.}(2008)Tinker, Kravtsov, Klypin, Abazajian, Warren,
  Yepes, Gottl{\"o}ber, \& Holz}]{tinker2008toward}
Tinker, J., Kravtsov, A.~V., Klypin, A., {et~al.} 2008, The Astrophysical
  Journal, 688, 709

\end{thebibliography}

\appendix
\section{3D cumulative galaxy density profiles}
\begin{figure*}[h!]
\centering
\includegraphics[width=1.1\linewidth,trim={4.5cm 1.4cm 1.5cm 0cm},clip]{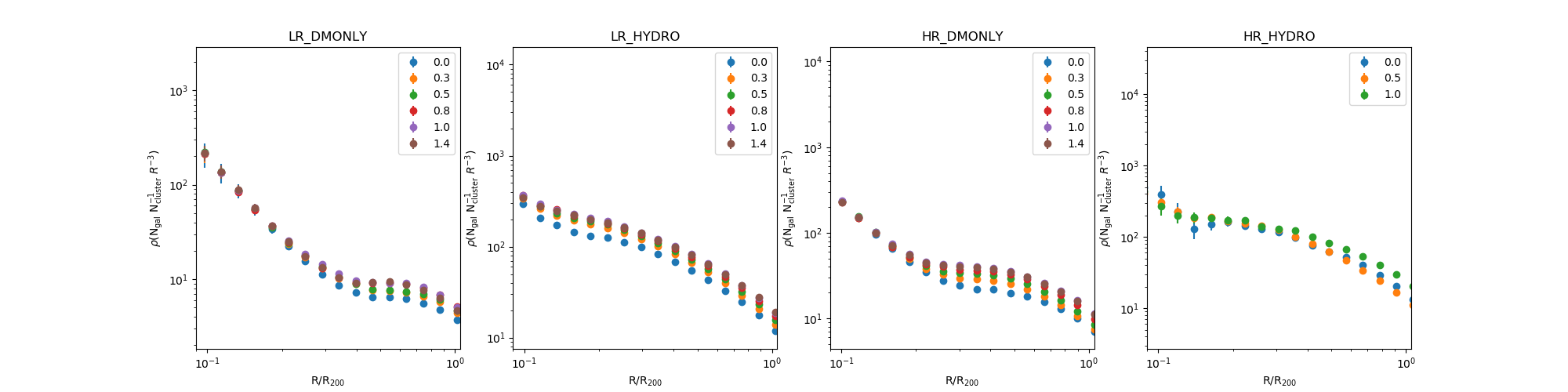}
\includegraphics[width=1.1\linewidth,trim={4.5cm 1.4cm 1.5cm 1.5cm },clip]{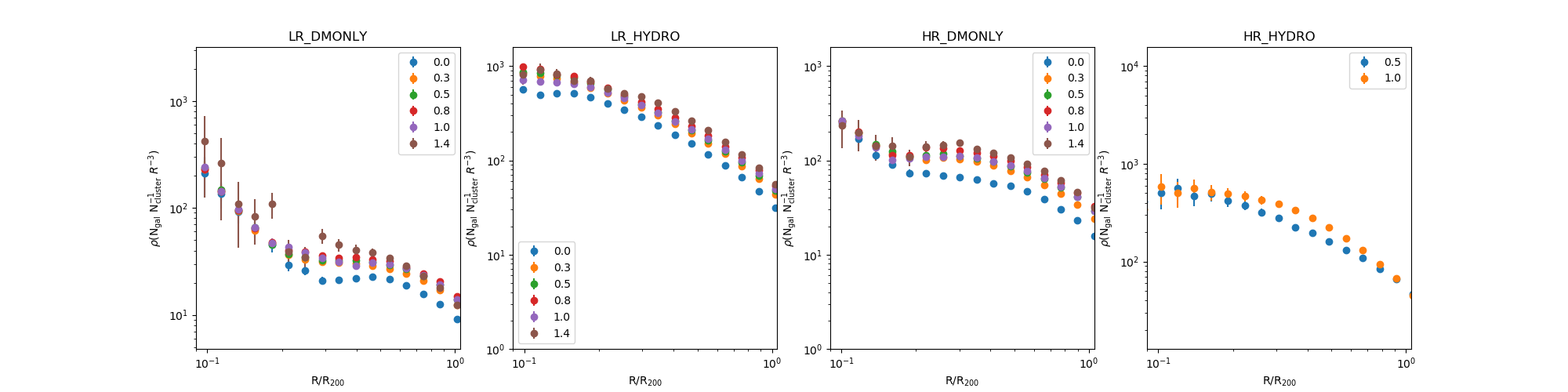}
\includegraphics[width=1.1\linewidth,trim={4.5cm 1.4cm 1.5cm 1.5cm},clip]{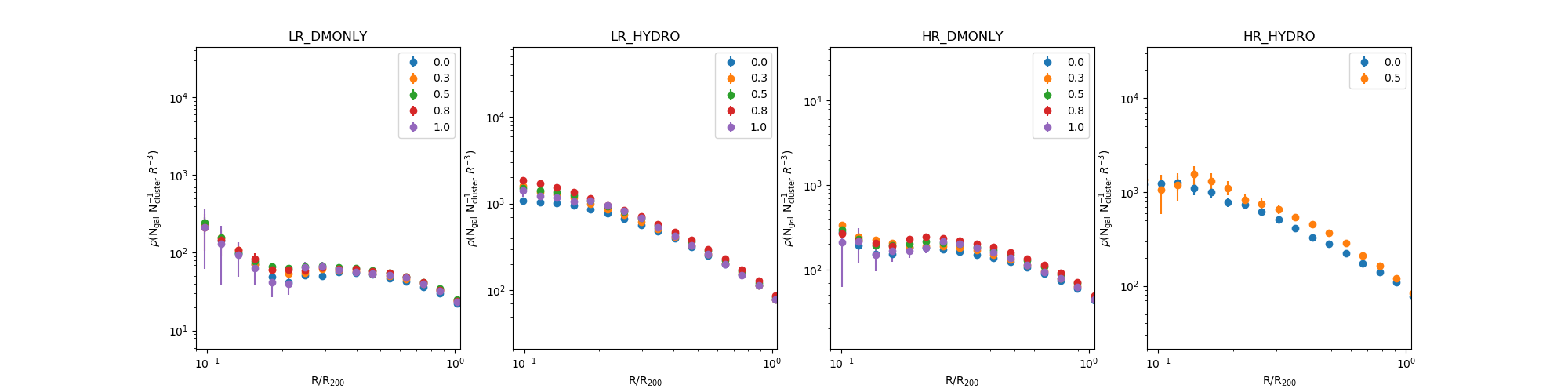}
\includegraphics[width=1.1\linewidth,trim={4.5cm 0 1.5cm 1.5cm},clip]{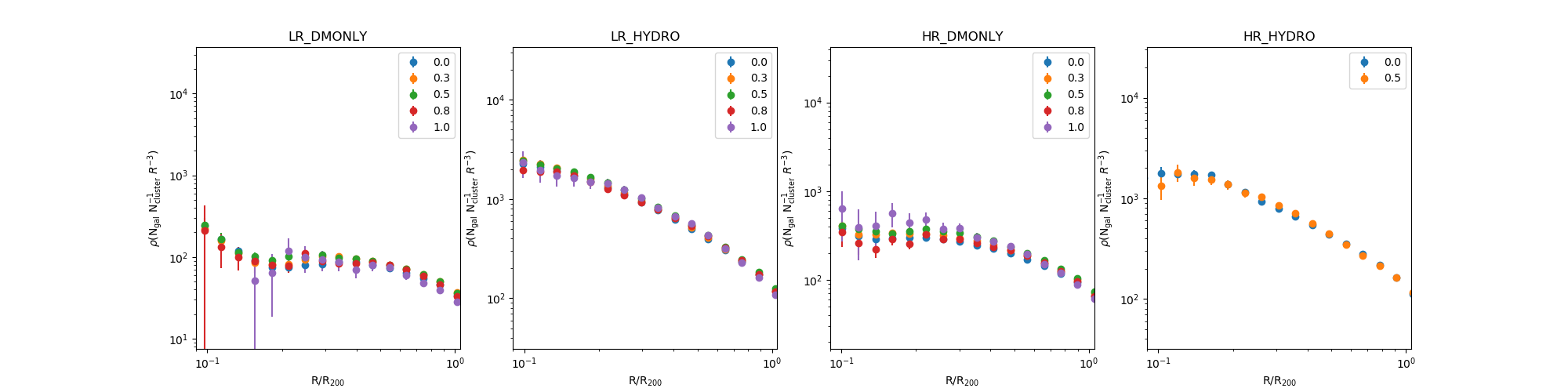}
\caption{Redshift evolution of 3D cumulative galaxy density profiles for the four simulation flavours and for the four bin in mass in Table~\ref{table_ch4:mass_bins}.  \label{fig:radialdistributionallz}}
\end{figure*}

We present in Figure~\ref{fig:radialdistributionallz} the 3D cumulative galaxy density distribution for the four simulation flavors considered -- \LRD, \LRH, \HRD\ and \HRH (from left to right). From top to bottom we present results for the four bins in mass presented in Table~\ref{table_ch4:mass_bins}. We plot the stacked density profiles for each slice in redshift. Notice that for the \HRH\ simulations we only dispose of five regions, which explains why profiles are missing at high redshift for the highest mass bins.

\end{document}